\documentclass[graybox]{svmult}

\usepackage{mathptmx}       
\usepackage{helvet}         
\usepackage{courier}        
\usepackage{type1cm}        
\usepackage{makeidx}         
\usepackage{graphicx}        
\usepackage{multicol}        
\usepackage[bottom]{footmisc}

\makeindex             

\begin{document}

\title*{Characterization of optical properties and surface roughness profiles:
The Casimir force between real materials}

\titlerunning{Characterization of optical properties}

\author{P. J. van Zwol, V.B. Svetovoy, and G. Palasantzas}
\institute{P. J. van Zwol \at Materials Innovation Institute and
Zernike Institute for Advanced Materials University of Groningen,
9747 AG Groningen, The Netherlands , \email{petervanzwol@gmail.com}
\and V. B. Svetovoy \at MESA$^+$ Institute for Nanotechnology,
University of Twente, PO 217, 7500 AE Enschede, The Netherlands
\email{v.b.svetovoy@ewi.utwente.nl} \and G. Palasantzas \at
Materials Innovation Institute and Zernike Institute for Advanced
Materials University of Groningen, 9747 AG Groningen, The
Netherlands \email{g.palasantzas@rug.nl}}
%
%
\maketitle


\abstract{The Lifshitz theory provides a method to calculate the
Casimir force between two flat plates if the frequency dependent
dielectric function of the plates is known. In reality any plate is
rough and its optical properties are known only to some degree. For
high precision experiments the plates must be carefully
characterized otherwise the experimental result cannot be compared
with the theory or with other experiments. In this chapter we
explain why optical properties of interacting materials are
important for the Casimir force, how they can be measured, and how
one can calculate the force using these properties. The surface
roughness can be characterized, for example, with the atomic force
microscope images. We introduce the main characteristics of a rough
surface that can be extracted from these images, and explain how one
can use them to calculate the roughness correction to the force. At
small separations this correction becomes large as our experiments
show. Finally we discuss the distance upon contact separating two
rough surfaces, and explain the importance of this parameter for
determination of the absolute separation between bodies.}

\section{Introduction}
\label{sec:1}

The Casimir force \cite{Cas48} between two perfectly reflecting
metals does not depend on the material properties. This is rather
rough approximation as the Lifshitz theory demonstrates
\cite{Lif55,DLP,LP9} (see also Dzyaloshinskii and Pitaevskii paper
in this volume). In this theory material dependence of the force
enters via the dielectric functions of the materials. Because the
Casimir-Lifshitz force originates from fluctuations of the
electromagnetic field, it is related with the absorption in the
materials via the fluctuation dissipation theorem. The dissipation
in the material at a frequency $\omega$ is proportional to the
imaginary part of the dielectric function
$\varepsilon(\omega)=\varepsilon^{'}(\omega)+i\varepsilon^{''}(\omega)
$. Thus, to predict the force one has to know the dielectric
properties of the materials.

In most of the experiments where the Casimir force was measured (see
reviews \cite{Lam05,Cap07} and Lamoreaux paper in this volume) the
bodies were covered with conducting films but the optical properties
of these films have never been measured. It is commonly accepted
that these properties can be taken from tabulated data in handbooks
\cite{HB1,Wea81}. Moreover, for conducting materials one has to know
also the Drude parameters, which are necessary to extrapolate the
data to low frequencies \cite{Lam00}. This might still be a possible
way to estimate the force, but it is unacceptable  for calculations
with controlled precision. The reason is very simple
\cite{Lam99b,Lam99a,Sve00a,Sve00b}: optical properties of deposited
films depend on the method of preparation, and can differ
substantially from sample to sample.

Analysis of existing optical data for Au \cite{Pir06} revealed
appreciable variation of the force in dependence on the optical data
used for calculations. Additionally, we measured our gold films
using ellipsomety in a wide range of wavelengths $0.14-33\;\mu$m
\cite{Sve08}, and found significant variation of optical properties
for samples prepared at different conditions. Considerable
dependence of the force on the precise dielectric functions of the
involved materials was also stressed for the system
solid-liquid-solid \cite{Zwo09a}.

The Lifshitz formula can be applied to two parallel plates separated
by a gap $d$. In reality each plate is rough and the formula cannot
be applied directly. When the separation of the plates is much
larger than their root-mean-square (rms) roughness $w$ one can
calculate correction to the force due to roughness using the
perturbation theory. But even in this case the problem is far from
trivial \cite{Gen03,Mai05a,Mai05b}. The roughness correction can be
easily calculated only if one can apply Proximity Force
Approximation (PFA) \cite{Der56}. Application of this approximation
to the surface profile is justified when this profile changes slowly
in comparison with the distance between bodies. Typical lateral size
of a rough body is given by the correlation length $\xi$. Then the
condition of applicability of PFA is $\xi\gg d$. This is very
restrictive condition since, for example, for thermally evaporated
metallic films the typical correlation length is $\xi\sim 50$ nm.

The roughness of the interacting bodies restricts the minimal
separation $d_0$ between the bodies. This distance (distance upon
contact) has a special significance for adhesion, which under dry
conditions is mainly due to Casimir/van der Waals forces across an
extensive noncontact area \cite{Del05}. It is important for micro
(nano) electro mechanical systems (MEMS) because stiction due to
adhesion is the major failure mode in MEMS \cite{Mab97}.
Furthermore, the distance upon contact plays an important role in
contact mechanics, is very significant for heat transfer, contact
resistivity, lubrication, and sealing.

Naively one could estimate this distance as the sum of the rms
roughnesses of body 1 and body 2, $d_0\approx w_1+w_2$ \cite{Hou97},
however, the actual minimal separation is considerably larger. This
is because $d_0$ is determined by the highest asperities rather then
those with the rms height. An empirical rule \cite{Zwo08b} for gold
films gives $d_0\approx 3.7\times (w_1+w_2)$ for the contact area of
$\sim\; 1\mu$m$^2$. The actual value of $d_0$ is a function of the
size of the contact area $L$. This is because for larger area the
probability to find a very high peak on the surface is larger.

Scale dependence (dependence on the size $L$) is also important for
the Casimir force in the noncontact regime. In this case there is an
uncertainty in the separation $\delta d(L)$, which depends on the
scale $L$. The reason for this uncertainty is the local variation of
the zero levels, which define the mathematical (average) surfaces of
the bodies. This uncertainty depends on the roughness of interacting
bodies and disappears in the limit $L\rightarrow \infty$.

In this paper we explain how one can collect the information about
optical properties of the materials, which is necessary for the
calculation of the Casimir-Lifshitz force. It is also discussed how
the optical spectra of different materials manifest themselves in
the force. We introduce the main characteristics of rough surfaces
and discuss how they are related with the calculation of the
roughness correction to the force. Scale dependence of the distance
upon contact is discussed, and we explain significance of this
dependence for the precise measurements of the force.

\section{Optical properties of materials and the Casimir force}
\label{sec:2}

Most Casimir force measurements were performed between metals
\cite{Lam05,Cap07,Lam97,Har00,Dec07} either e-beam evaporated or
plasma sputtered on substrates. For such metallic films the grains
are rather small in the order of tens of nanometers, and the amount
of defects and voids is large \cite{Asp80}. The force measured
between silicon single crystal and gold coated sphere \cite{Che05}
simplify situation only partly: the optical properties of the
Si-crystal are well defined but properties of Au coating are not
known well.

A detailed literature survey performed by Pirozhenko et al.
\cite{Pir06} revealed significant scatter in the dielectric data of
gold films collected by different groups. The measurement errors
were not large and could not explain the data scattering. It was
concluded that scattering of the data for gold films could lead to
uncertainty in the calculated force up to 8\% at separations around
100 nm. Most of the optical data for metals do not extend beyond the
wavelength of $14\;\mu\textrm{m}$ in the infrared range
\cite{Dol65,Mot64}. Thus, it would be important to explore more the
infrared regime and compare modern measured optical properties of
samples used in force measurements with the old data. Moreover, mid
and far infrared data are very important for the force prediction
(see Sec. \ref{sec:2.2.1}). This was accomplished by Svetovoy et al.
\cite{Sve08} where ellipsometry from the far infrared (IR) to near
ultraviolet (UV) was used over the wavelength range
$140\;\textrm{nm}-33\;\mu\textrm{m}$ to obtain the frequency
dependent dielectric functions for gold films prepared in different
conditions. Analysis of different literature sources where the
dielectric functions of a number of dielectrics such as silica and
some liquids was performed by van Zwol et al. \cite{Zwo09a}.
Situations where the data scattering can change even the qualitative
behavior of the force (from attractive to repulsive) were indicated.

\subsection{Dielectric function in the Casimir force}
\label{sec:2.1}

In this section we discuss how the dielectric functions of
materials enter the Lifshitz theory and how these functions can be
found experimentally.

\subsubsection{The force}
\label{sec:2.1.1}

Let us start the discussion from the Lifshitz formula \cite{LP9}
between two parallel plates separated by a gap $d$. It has the
following form
\begin{equation}\label{Lif_imag}
    F(T,d)=\frac{kT}{\pi}{\sum\limits_{n=0}^{\infty}}\;^{'}
    \int\limits_0^{\infty}dqq\kappa_0\sum_{\nu=s,p}
    \frac{r^{\nu}_{1}r^{\nu}_{2}e^{-2\kappa_0 d}}
    {1-r^{\nu}_{1}r^{\nu}_{2}e^{-2\kappa_0 d}},
\end{equation}
where "prime" at the sign of sum means that the $n=0$ term must be
taken with the weight 1/2, the wave vector in the gap is
$\textbf{K}=(\textbf{q},\kappa_0)$ with the $z$-component
$\kappa_0$ defined below. Index "0" is related with the gap. Here
$r^{\nu}_{1,2}$ are the reflection coefficients of the inner
surfaces of the plates (index 1 or 2) for two different
polarizations: $\nu=s$ or transverse electric (TE) polarization,
and $\nu=p$ or transverse magnetic (TM) polarization. Specific of
the Lifshitz formula in the form (\ref{Lif_imag}) is that it is
defined for a discrete set of imaginary frequencies called the
Matsubara frequencies
\begin{equation}\label{Matsubara}
    \omega_n=i\zeta_n=i\frac{2\pi kT}{\hbar}n,\ \ \ n=0,1,2,...,
\end{equation}
where $T$ is the temperature of the system and $n$ is the
summation index.

In practice the interacting bodies are some substrates covered with
one or a few layers of working materials. If the top layer can be
considered as a bulk layer then the reflection coefficients for body
$i$ are given by simple Fresnel formulas \cite{LL8}:
\begin{equation}\label{refl}
    r^s_i=\frac{\kappa_0-\kappa_i}{\kappa_0+\kappa_i}, \ \ \
    r^p_i=\frac{\varepsilon_i(i\zeta)\kappa_0-\varepsilon_0(i\zeta)\kappa_i}
    {\varepsilon_i(i\zeta)\kappa_0+\varepsilon_0(i\zeta)\kappa_i},
\end{equation}
where
\begin{equation}\label{ki_def}
    \kappa_0=\sqrt{\varepsilon_0(i\zeta)\frac{\zeta^2}{c^2}+q^2}, \ \ \
    \kappa_i=\sqrt{\varepsilon_i(i\zeta)\frac{\zeta^2}{c^2}+q^2}.
\end{equation}
For multilayered bodies these formulas can be easily generalized
(in relation with the dispersive forces see Ref. \cite{Zho95}).
Only the reflection coefficients depend on the dielectric
functions of the plate materials; the function
$\varepsilon_0(i\zeta)$ of the gap material enters additionally in
$\kappa_0$.

At small separations the thermal dependence of the force is very
weak and in many cases can be neglected. Because important imaginary
frequencies are around the so called characteristic frequency
$\zeta_c=c/2d$, then the relative thermal correction can be
estimated as $kT/\hbar\zeta_c$. For room temperature
$T=300^{\circ}\;\textrm{K}$ and separations smaller than
$100\;\textrm{nm}$ the correction will be smaller than 3\%. If one
can neglect this correction then in the Lifshitz formula $\zeta$ can
be considered as a continuous variable and the sum in
(\ref{Lif_imag}) is changed by the integral according to the rule:
\begin{equation}\label{subs}
    \frac{kT}{\pi}{\sum\limits_{n=0}^{\infty}}\;^{'}\rightarrow
    \frac{\hbar}{2\pi^2}\int\limits_0^{\infty}d\zeta.
\end{equation}

The material function $\varepsilon(i\zeta)$ (we suppress the indexes
for a while) cannot be measured directly but can be expressed via
the observable function $\varepsilon^{''}(\omega)$ with the help of
the Kramers-Kronig (KK) relation \cite{LL8}:
\begin{equation}\label{KK_imag}
    \varepsilon(i\zeta)=1+\frac{2}{\pi}\int\limits_0^{\infty}d\omega
    \frac{\omega\varepsilon^{''}(\omega)}{\omega^2+\zeta^2}.
\end{equation}
The knowledge of $\varepsilon(i\zeta)$ is of critical importance for
the force calculations. Equation (\ref{KK_imag}) demonstrates the
main practical problem. To find the function $\varepsilon(i\zeta)$
for $\zeta\sim \zeta_c$ in general one has to know the physical
function $\varepsilon^{''}(\omega)$ in a wide range of real
frequencies, which is not necessary coincides with $\omega\sim
\zeta_c$. This property of the Casimir force was stressed in Ref.
\cite{Sve00a} and then was demonstrated experimentally
\cite{Ian04,Lis05,Man09}. It will be discussed below on specific
examples.

\subsubsection{The optical data}
\label{sec:2.1.2}

The optical properties of materials are described by two measurable
quantities: the index of refraction $n(\lambda)$ and the extinction
coefficient $k(\lambda)$, which both depend on the wavelength of the
electromagnetic field $\lambda$. Combined together they define the
complex index of refraction
$\tilde{n}(\lambda)=n(\lambda)+ik(\lambda)$. The real part defines
the phase velocity in a medium $v=c/n$ where $c$ is the speed of
light. The imaginary part tells us how much light is adsorbed when
it travels through the medium. The dielectric response of a material
for the UV ($\hbar\omega>5\; \textrm{eV}$), IR ($0.01-1\;
\textrm{eV}$) and MicroWave (MW) or TeraHertz range
($10^{-4}-10^{-2}\; \textrm{eV}$), is related to electronic
polarization resonances, atomic polarization resonances (in case of
metals this is a gas of quasi free electrons), and dipole
relaxation, respectively.

The complex dielectric function
$\varepsilon(\omega)=\varepsilon^{\prime}(\omega)+
i\varepsilon^{\prime\prime}(\omega)$
and the complex index of refraction are related as
$\varepsilon(\omega)=\tilde{n}^2(\omega)$, which is equivalent to
the following equations:
\begin{equation}\label{eps_n}
    \varepsilon^{'}=n^2-k^2,\ \ \ \varepsilon^{''}=2nk.
\end{equation}
In many cases only the absorbance is measured for a given material.
In this case the refraction index can be found from the KK relation
at real frequencies:
\begin{equation}\label{KK_eps}
    \varepsilon^{'}(\omega)=1+\frac{2}{\pi}P\int\limits_0^{\infty}
    dx\frac{x\varepsilon^{''}(x)}{x^2-\omega^2},
\end{equation}
where $P$ means the principal part of the integral.

Kramers-Kronig relations originating from causality have a very
general character. They are useful in dealing with experimental
data, but one should be careful since in most cases dielectric data
are available over limited frequency intervals. As a result specific
assumptions must be made about the form of the dielectric data
outside of measurement intervals, or the data should be combined
with other (tabulated) experimental data before performing the KK
integrals.

A powerful method to collect optical data simultaneously for both
$\varepsilon^{'}$ and $\varepsilon^{''}$ is ellipsomery. This is a
non destructive technique where one measures an intensity ratio
between incoming and reflected light and the change of the
polarization state. Ellipsometry is less affected by intensity
instabilities of the light source or atmospheric absorption. Because
the ratio is measured no reference measurement is necessary. Another
advantage is that both real and imaginary parts of the dielectric
function can be extracted without the necessity to perform a
Kramers-Kronig analysis. The ellipsometry measures two parameters
$\Psi$ and $\Delta$, which can be related to the ratio of complex
Fresnel reflection coefficients for p- and s-polarized light
\cite{Azz87,Tom99}
\begin{equation}\label{rho_def}
    \rho=\frac{r^p}{r^s}=\tan\Psi e^{i\Delta},
\end{equation}
where $r^{p,s}$ are the reflection coefficients of the investigated
surface, and the angles $\Psi$ and $\Delta$ are the raw data
collected in a measurement as functions of the wavelength $\lambda$.
When the films are completely opaque (bulk material), then the
reflection coefficients are related with the dielectric function as
follows
\begin{equation}
r_p=\frac{\left\langle\varepsilon\right\rangle\cos\vartheta-
\sqrt{\left\langle\varepsilon\right\rangle-\sin^2\vartheta}}
{\left\langle\varepsilon\right\rangle\cos\vartheta+
\sqrt{\left\langle\varepsilon\right\rangle-\sin^2\vartheta}},\ \ \
r_s=\frac{\cos\vartheta-\sqrt{\left\langle\varepsilon\right\rangle-\sin^2\vartheta}}
{\cos\vartheta+\sqrt{\left\langle\varepsilon\right\rangle-\sin^2\vartheta}},
\label{refl}
\end{equation}
where $\vartheta$ is the angle of incidence and
$\left\langle\varepsilon\right\rangle=\left\langle\varepsilon(\lambda)\right\rangle$
is the "pseudo" dielectric function of the films. The term "pseudo"
is used here since the films may not be completely isotropic or
uniform; they are rough, and may contain absorbed layers of
different origin because they have been exposed to air. If this is
the case then the dielectric function extracted from the raw data
will be influenced by these factors. The dielectric function is
connected with the ellipsometric parameter $\rho$ for an isotropic
and uniform solid as
\begin{equation}
\varepsilon=\sin^2\vartheta\left[1+\tan^2
\vartheta\left(\frac{1-\rho}{1+\rho}\right)^2\right].
\label{Ell_eps}
\end{equation}

As it was stated before the spectral range of our measured data is
from $137\; \textrm{nm}$ to $33\;\mu\textrm{m}$. Even longer
wavelengths have to be explored to predict the force between
metals without using the extrapolation. Ellipsometry in the
terahertz range $0.1-8\;\textrm{THz}$ (wavelengths
$38-3000\;\mu\textrm{m}$) is difficult due to lack of intense
sources in that range, and these systems are still in development
\cite{Hof09}. Typically synchrotron radiation is used as a source
deeming these measurements very expensive. Nonetheless for gold
films it would be extremely interesting to have dielectric data in
this regime.

Dielectric data obtained by ellipsometry or absorption measurements
\cite{Fen02} in the far UV regime are also rare. The most obvious
reason for this is that these measurements are expensive because
high energy photons must be produced, again at synchrotrons
\cite{Tom05}. Furthermore, ellipsometry in this range becomes
complicated as polarizing materials become non transparent. For this
range a few ellipsometry setups exist around the world covering the
range $5-90\; \textrm{eV}$ (wavelengths $12-200\; \textrm{nm}$)
\cite{Tom05}. The vacuum UV (VUV) and extreme UV (XUV) parts may not
be very important for metals but for low permittivity dielectrics
such as all liquids, and, for example, silica or teflon, there is a
major absorption band in the range $5-100\; \textrm{eV}$ (see Fig.
\ref{fig6}). It is precisely this band that dominates in the
calculations of the Casimir force for these materials. It is very
unfortunate that precisely for this band dielectric data are lacking
for most substances except for a few well know cases as, for
example, water.

\subsection{Gold films}
\label{sec:2.2}

In this section we discuss optical characterization of our gold
films prepared in different conditions using ellipsometers.  Then we
discuss the dielectric function at imaginary frequencies for Au
films and for metals in general stressing the importance of very low
real frequencies for precise evaluation of $\varepsilon(i\zeta)$ at
$\zeta\sim\zeta_c\sim 1\;\textrm{eV}$ (separations around
$100\;\textrm{nm}$). Finally we describe variation in the Casimir
force if different samples would be used for the force measurements.

\subsubsection{$\varepsilon(\omega)$ for Au films}
\label{sec:2.2.2}

\begin{figure}[tbp]
\includegraphics[width=11.5cm]{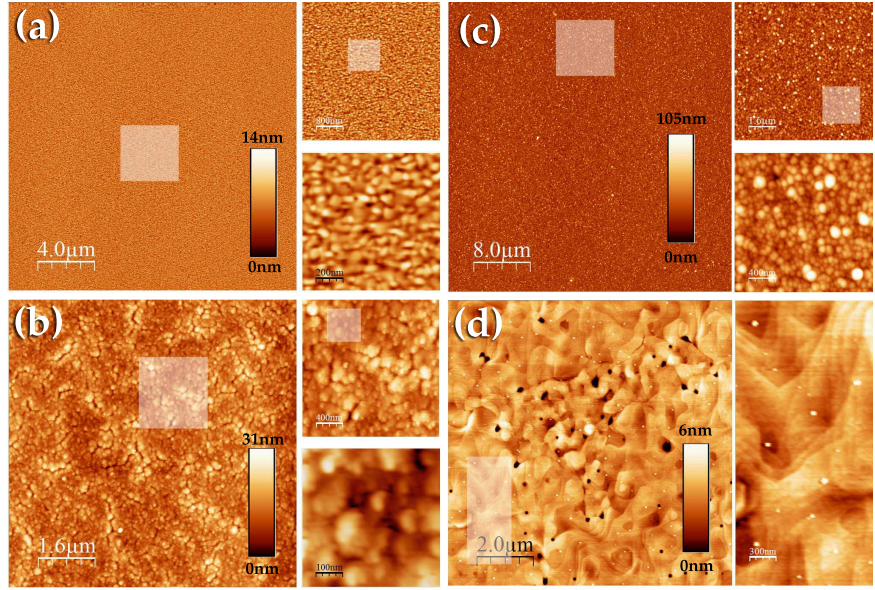}
\caption{Flattened roughness scans (up to $4000\times 4000$ pixels)
of gold surfaces where the highlighted areas are magnified. The
color scale bars can be applied only to the large images. (a) $100\;
\textrm{nm}$ Au on Si, (b) Au coated polysterene sphere (first
plasma sputtered then $100\; \textrm{nm}$ Au evaporated, (c) $1600\;
\textrm{nm}$ Au on Si, (d) very high quality $120\; \textrm{nm}$ Au
on mica, annealed for a few hours and slowly cooled down. }
\label{fig1}
\end{figure}

Let us have now a closer look at the dielectric functions of our
gold films \cite{Sve08} used for the force measurements in Refs.
\cite{Zwo08a,Zwo08b}. For optical characterization we have
prepared five films by electron beam evaporation. Three of these
films of different thicknesses 100, 200 and $400\; \textrm{nm}$
were prepared within the same evaporation system on Si with $10\;
\textrm{nm}$ titanium sublayer and were not annealed. Different
evaporation system was used to prepare two other films. These
films were $120\; \textrm{nm}$ thick. One film was deposited on
mica and was extensively annealed. The other one was deposited on
Si with chromium sublayer and was not annealed.

The AFM scans of the $100\; \textrm{nm}$ film and the annealed
$120\; \textrm{nm}$ film on mica are shown in Fig. \ref{fig1}. In
the same figure are shown also the gold covered sphere and $1600\;
\textrm{nm}$ film, which where not used in optical characterization.
One can see that the annealed sample is atomically smooth over
various length scales with atomic steps and terraces visible.
Nevertheless, the local trenches of $5\; \textrm{nm}$ deep are still
present.

Optical characterization of the films was performed by J. A. Woollam
Co., Inc. \cite{Woo}. Vacuum ultraviolet variable angle
spectroscopic ellipsometer (VASE) was used in the spectral range
$137-1698\;\textrm{nm}$. In the spectral range
$1.9-32.8\;\mu\textrm{m}$ the infrared variable angle spectroscopic
ellipsometer (IR-VASE) was used for two incidence angle of
$65^{\circ}$ and $75^{\circ}$. The roughness and possible absorbed
layer on the film surface can have some significance in the visible
and ultraviolet ranges but not in the infrared, where the absorption
on free electrons of metals is very large. Moreover, the effect of
roughness is expected to be small since for all films the rms
roughness is much smaller than the smallest wavelength
$137\;\textrm{nm}$. Because the infrared domain is the most
important for the Casimir force between metals, we will consider
$\left\langle\varepsilon(\lambda)\right\rangle$ extracted from the
raw data as a good approximation for the dielectric function of a
given gold film.

\begin{figure}[tbp]
\includegraphics[width=5.75cm]{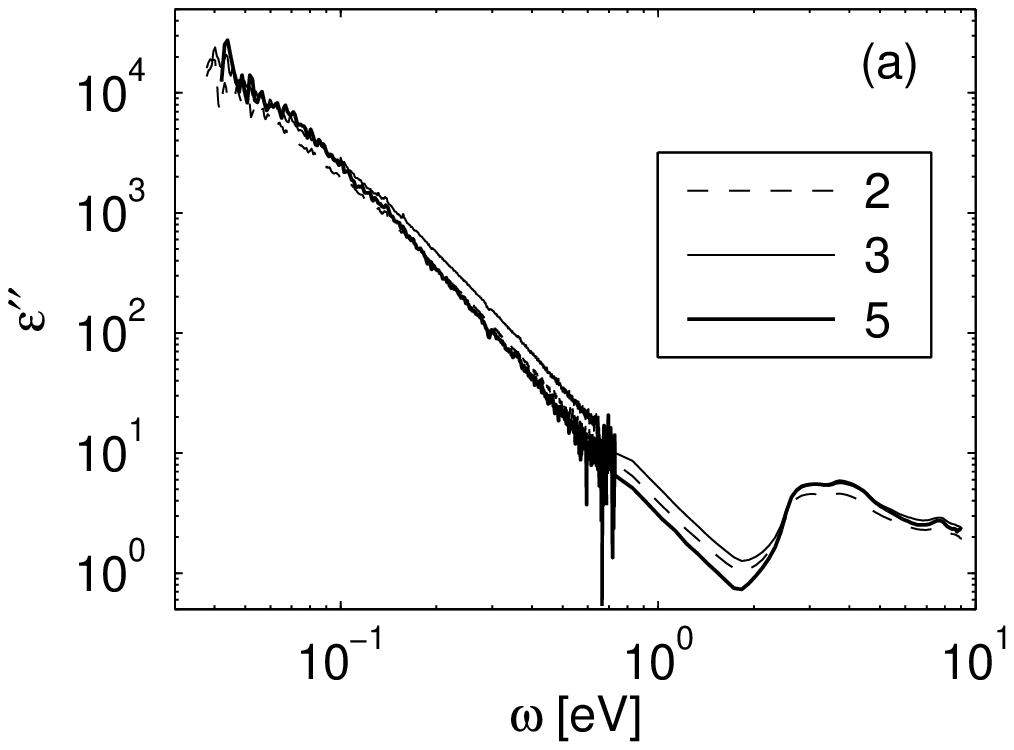}
\includegraphics[width=5.75cm]{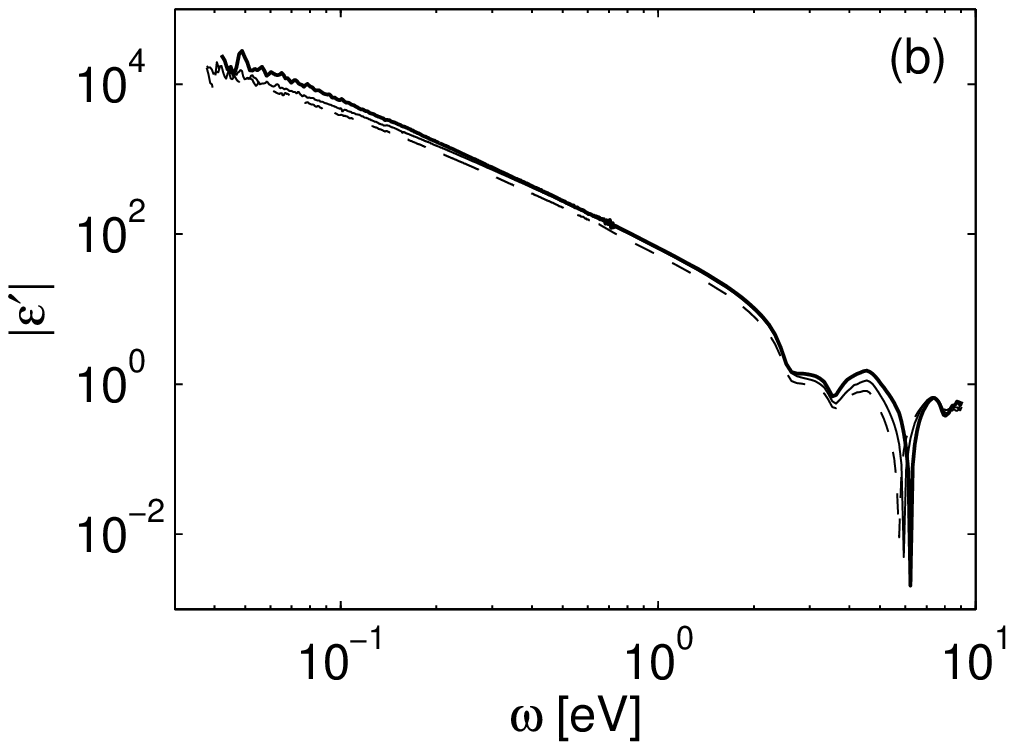}\newline
\caption{(a) Measured $\varepsilon^{\prime\prime}$ as a function of
frequency $\omega$. (b) The same for $|\varepsilon^{\prime}|$. For
clearness the results are presented only for three samples 2, 3, and
5. } \label{fig2}
\end{figure}

Figure \ref{fig2}(a) shows the experimental results for
$\varepsilon^{''}(\omega)$ for three of five investigated samples.
Around the interband transition (minimum of the the curves) the
smallest absorption is observed for the sample 5 (annealed on mica)
indicating the smallest number of defects in this sample
\cite{Asp80}. On the contrary, this sample shows the largest
$|\varepsilon^{'}(\omega)|$ in the infrared as one can see in Fig.
\ref{fig2}b. An important conclusion that can be drawn from our
measurements is the sample dependence of the dielectric function.
The sample dependence can be partly attributed to different volumes
of voids in films as was proposed by Aspnes {\it et al.}
\cite{Asp80}. The values of $\varepsilon$ and their dispersion for
different samples are in good correspondence with old measurements
\cite{Dol65,Mot64}. The log-log scale is not very convenient for
having an impression of this dependence. We present in Table I the
values of $\varepsilon$ for all five samples at chosen wavelengths
$\lambda=1,5,10\;\mu\textrm{m}$. One can see that the real part of
$\varepsilon$ varies very significantly from sample to sample.

\begin{table}
\centering
\begin{tabular}{p{3.0cm}p{2.5cm}p{2.5cm}p{3.0cm}}
\hline\noalign{\smallskip}
Sample & $\lambda=1\;\mu\textrm{m}$ & $\lambda=5\;\mu\textrm{m}$ & $\lambda=10\;\mu\textrm{m}$ \\
\hline\hline 1, 400 nm/Si & $-29.7+i 2.1$ & $-805.9+i 185.4$ & $-2605.1+i 1096.3$ \\
   2, 200 nm/Si & $-31.9+i 2.3$ & $-855.9+i 195.8$ & $-2778.6+i 1212.0$ \\
   3, 100 nm/Si & $-39.1+i 2.9$ & $-1025.2+i 264.8$ & $-3349.0+i 1574.8$ \\
   4, 120 nm/Si & $-43.8+i 2.6$ & $-1166.9+i 213.9$ & $-3957.2+i 1500.1$ \\
   5, 120 nm/mica & $-40.7+i 1.7$ & $-1120.2+i 178.1$ & $-4085.4+i
   1440.3$\\
\noalign{\smallskip}\hline\noalign{\smallskip}
\end{tabular}
\caption{Dielectric function for different samples at fixed
wavelengths $\lambda=1,5,10\;\mu\textrm{m}$.}\label{tab1}
\end{table}

One could object that the real part of $\varepsilon$ does not play
role for $\varepsilon(i\zeta)$ and only variation of
$\varepsilon^{''}(\omega)$ from sample to sample is important.
However, both $\varepsilon^{''}(\omega)$ and
$\varepsilon^{'}(\omega)$ are important for precise determination of
the Drude parameters. Let us discuss now how one can extract these
parameters from the data.

All metals have finite conductivity. It means that at low
frequencies $\omega\rightarrow 0$ the dielectric function behaves as
$\varepsilon(\omega)\rightarrow 4\pi\sigma/\omega$, where $\sigma$
is the material conductivity. It has to be stressed that this
behavior is a direct consequence of the Ohm's law and, therefore, it
has a fundamental character. Because the dielectric function has a
pole at $\omega\rightarrow 0$, the low frequencies will give a
considerable contribution to $\varepsilon(i\zeta)$ even if $\zeta$
is high (for example, in visible part of the spectrum) as one can
see from Eq.(\ref{KK_imag}).

Usually it is assumed that at low frequencies the dielectric
functions of good metals follow the Drude model:
\begin{equation}\label{Drude}
    \varepsilon(\omega)=1-\frac{\omega_p^2}{\omega\left(\omega+i\gamma\right)},
\end{equation}
where $\omega_p$ is the plasma frequency and $\gamma$ is the
relaxation frequency of a given metal. When $\omega\rightarrow 0$ we
reproduce the $1/\omega$ behavior with the conductivity
$\sigma=\omega_p^2/4\pi\gamma$. For good metals such as Au, Ag, Cu,
Al typical values of the Drude parameters are $\omega_p\sim
10^{16}\;\textrm{rad/s}$ and $\gamma\sim 10^{14}\;\textrm{rad/s}$.

Separating real and imaginary parts in Eq. (\ref{Drude}) one finds
for $\varepsilon^{'}$ and $\varepsilon^{''}$
\begin{equation}\label{Drude_RI}
    \varepsilon^{'}(\omega)=1-\frac{\omega_p^2}{\omega^2+\gamma^2},\
    \ \
    \varepsilon^{''}(\omega)=\frac{\omega_p^2\gamma}{\omega\left(
    \omega^2+\gamma^2\right)}.
\end{equation}
These equations can be applied below the interband transition
$\omega<2.45\;\textrm{eV}$ ($\lambda>0.5\;\mu\textrm{m}$)
\cite{The70}, but this transition is not sharp and one has to do
analysis at lower frequencies. Practically Eq. (\ref{Drude_RI}) can
be applied at wavelengths $\lambda>2\;\mu\textrm{m}$ that coincides
with the range of the infrared ellipsometer. The simplest way to
find the Drude parameters is to fit the experimental data for
$\varepsilon^{'}$ and $\varepsilon^{''}$ with both equations
(\ref{Drude_RI}). Alternatively to find the Drude parameters one can
use the functions $n(\omega)$ and $k(\omega)$ and their Drude
behavior, which follows from the relation
$\varepsilon(\omega)=\tilde{n}^2(\omega)$. This approach uses
actually the same data but weight noise differently.

Completely different but more complicated approach is based on the
KK relations (\ref{KK_eps}) (see Refs. \cite{Pir06} and \cite{Sve08}
for details). In this case one uses measured
$\varepsilon^{''}(\omega)$ extrapolated to low frequencies according
to the Drude model and extrapolated to high frequencies as $
A/\omega^3$, where $A$ is a constant. In this way we will get
$\varepsilon^{''}(\omega)$ at all frequencies. Using then Eq.
(\ref{KK_eps}) we can predict $\varepsilon^{'}(\omega)$. Comparing
the prediction with the measured function we can determine the Drude
parameters. Similar procedure can be done for $n(\omega)$ and
$k(\omega)$.

\begin{table}
\centering
\begin{tabular}{p{1.6cm}p{1.8cm}p{1.8cm}p{1.8cm}p{1.8cm}p{2cm}}
\hline\noalign{\smallskip}
  & 1, 400 nm/Si & 2, 200 nm/Si & 3, 100 nm/Si & 4, 120 nm/Si & 5, 120 nm/mica \\
\hline\hline $\gamma$ (meV) & $40.5\pm 2.1$ & $49.5\pm 4.4$ & $49.0\pm 2.1$ & $35.7\pm 5.1$ & $37.1\pm 1.9$ \\
  $\omega_p$ (eV) & $6.82\pm 0.08$ & $6.83\pm 0.15$ & $7.84\pm 0.07$ & $8.00\pm 0.16$ & $8.38\pm 0.08$ \\
  $\xi$ (nm) & 22 & 26 & 32 & 70 & 200 \\
  $w$ (nm) & 4.7 & 2.6 & 1.5 & 1.5 & 0.8 \\
\noalign{\smallskip}\hline\noalign{\smallskip}
\end{tabular}
\caption{Drude parameters $\gamma$, $\omega_p$ and roughness
parameters, the correlation length $\xi$ and rms roughness $w$, for
all five measured samples.}\label{tab2}
\end{table}

All methods for determination of the Drude parameters give
reasonably close values of both parameters. We cannot give
preference to any specific method. Instead, we average the values
of the parameters determined by different methods, and define the
rms error of this averaging as uncertainty in the parameter value.
The averaged parameters and rms errors are given in the Table
\ref{tab2}. We included in this table also the correlation lengths
$\xi$ and the rms roughness $w$ for the sample roughness profiles.
One can see the $\omega_p$ and $\xi$ correlate with each other
\footnote{This correlation was not noted in \cite{Sve08} and
stressed here for the first time.}. This correlation has sense
because $\xi$ describes the average size of the crystallites in
the film; the larger the crystallites the smaller number of the
defects has the film and, therefore, the larger value of the
plasma frequency is realized.

\begin{figure}[tbp]
\includegraphics[width=5.75cm]{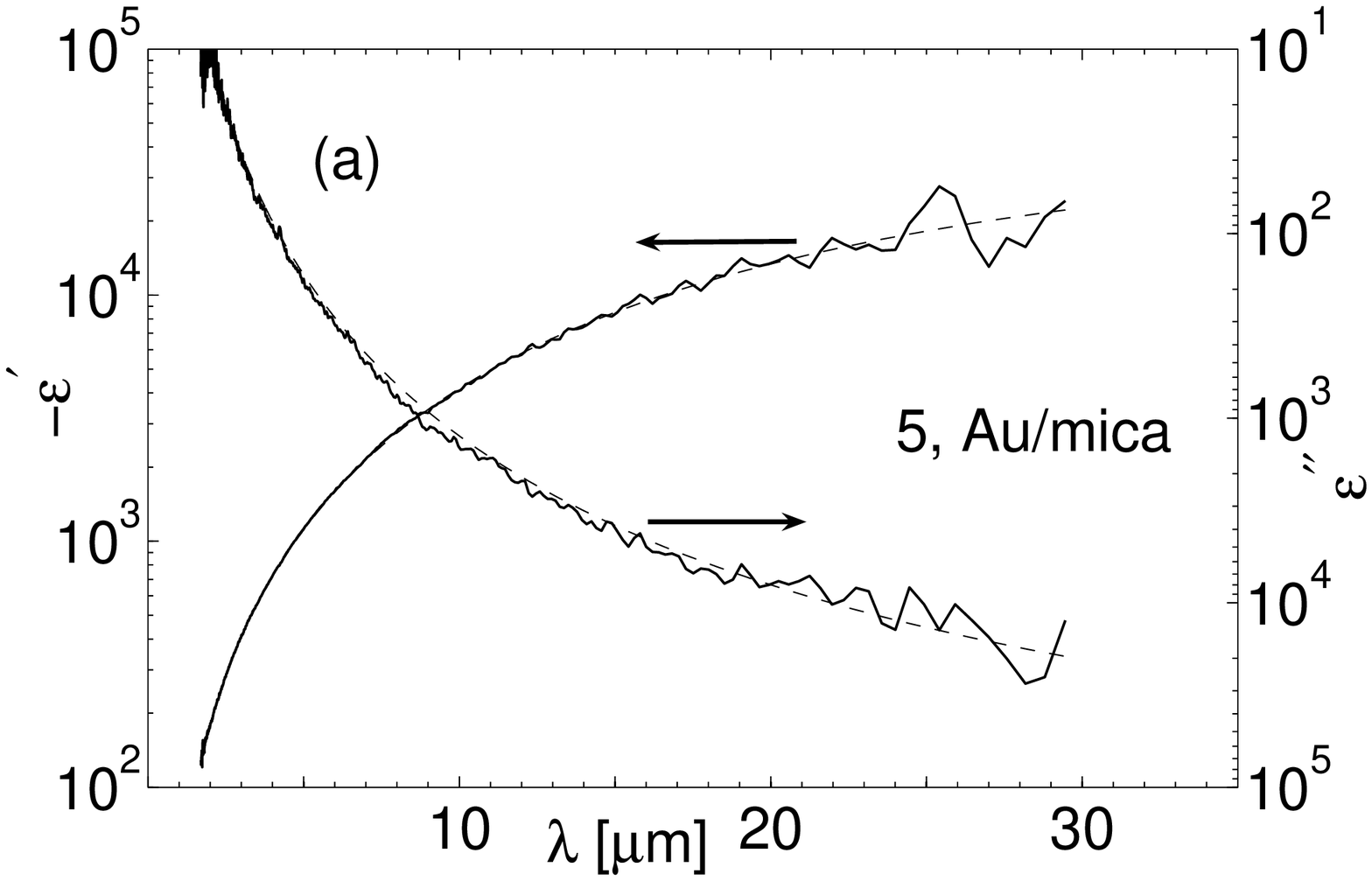}
\includegraphics[width=5.75cm]{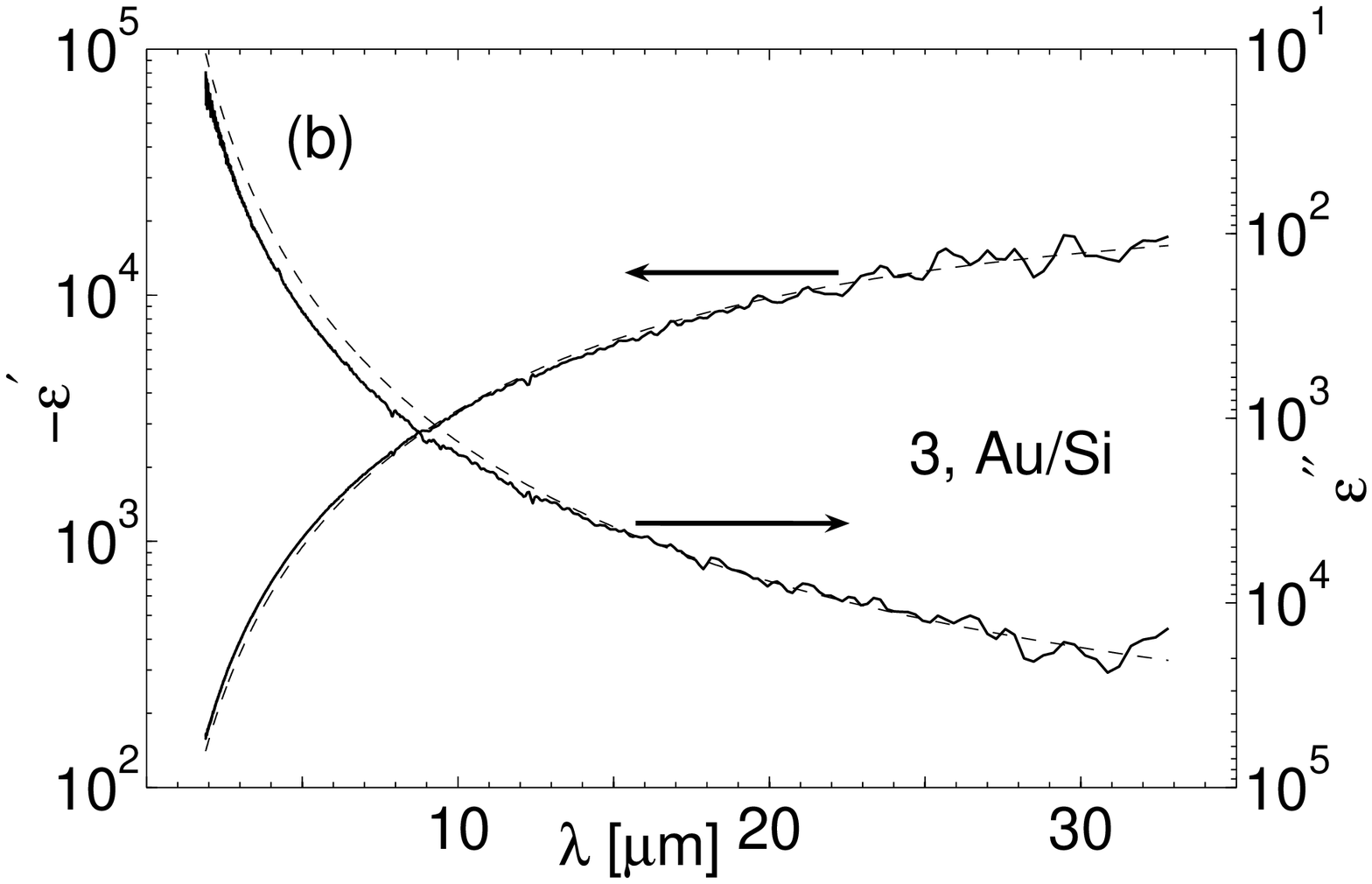}\newline
\caption{The infrared data as functions of the wavelength $\lambda$
for $\varepsilon^{'}$ and $\varepsilon^{''}$ (solid lines) and the
best Drude fits (dashed lines) for two gold films. Panel (a) shows
the data for annealed sample 5 and panel (b) shows the same for
unannealed sample 3.} \label{fig3}
\end{figure}

Quality of the Drude fit one can see in Fig. \ref{fig3} for samples
3 and 5. The fit is practically perfect for high quality sample 5,
but there are some deviations for sample 3 at short wavelengths
especially visible for $\varepsilon^{''}$. More detailed analysis
\cite{Sve08} revealed presence of a broad absorption peak of unknown
nature around $\lambda=10\;\mu\textrm{m}$. The magnitude of this
absorption is the largest for poor quality samples 1 and 2.

\subsubsection{$\varepsilon(i\zeta)$ for metals}
\label{sec:2.2.1}

Any method of optical characterization has a minimal accessible
frequency $\omega_{cut}$ (cut-off frequency). At
$\omega>\omega_{cut}$ one can measure the dielectric function but at
lower frequencies $\omega<\omega_{cut}$ one has to make an
assumption on the behavior of $\varepsilon(\omega)$, i. e.
extrapolate to low frequencies. In KK relation (\ref{KK_imag}) one
can separate two intervals: $\omega<\omega_{cut}$, where
$\varepsilon^{''}(\omega)$ has to be extrapolated and
$\omega>\omega_{cut}$, where $\varepsilon^{''}(\omega)$ is measured.
Then we can present $\varepsilon(i\zeta)$ as
\begin{eqnarray}\label{extr}
   \nonumber \varepsilon(i\zeta)=1+\varepsilon_{cut}(i\zeta)+\varepsilon_{exper}(i\zeta),
    \\
    \varepsilon_{cut}(i\zeta)=\frac{2}{\pi}\int\limits_0^{\omega_{cut}}
    d\omega\frac{\omega\varepsilon^{''}(\omega)}{\omega^2+\zeta^2},\
    \ \ \varepsilon_{exper}(i\zeta)=\frac{2}{\pi}\int\limits_{\omega_{cut}}^{\infty}
    d\omega\frac{\omega\varepsilon^{''}(\omega)}{\omega^2+\zeta^2}.
\end{eqnarray}
Of course, at very high frequencies we also do not know
$\varepsilon^{''}(\omega)$ but, for metals high frequencies are not
very important. For this reason we include the high frequency
contribution to $\varepsilon_{exper}(i\zeta)$.

\begin{figure}[tbp]
\includegraphics[width=5.75cm]{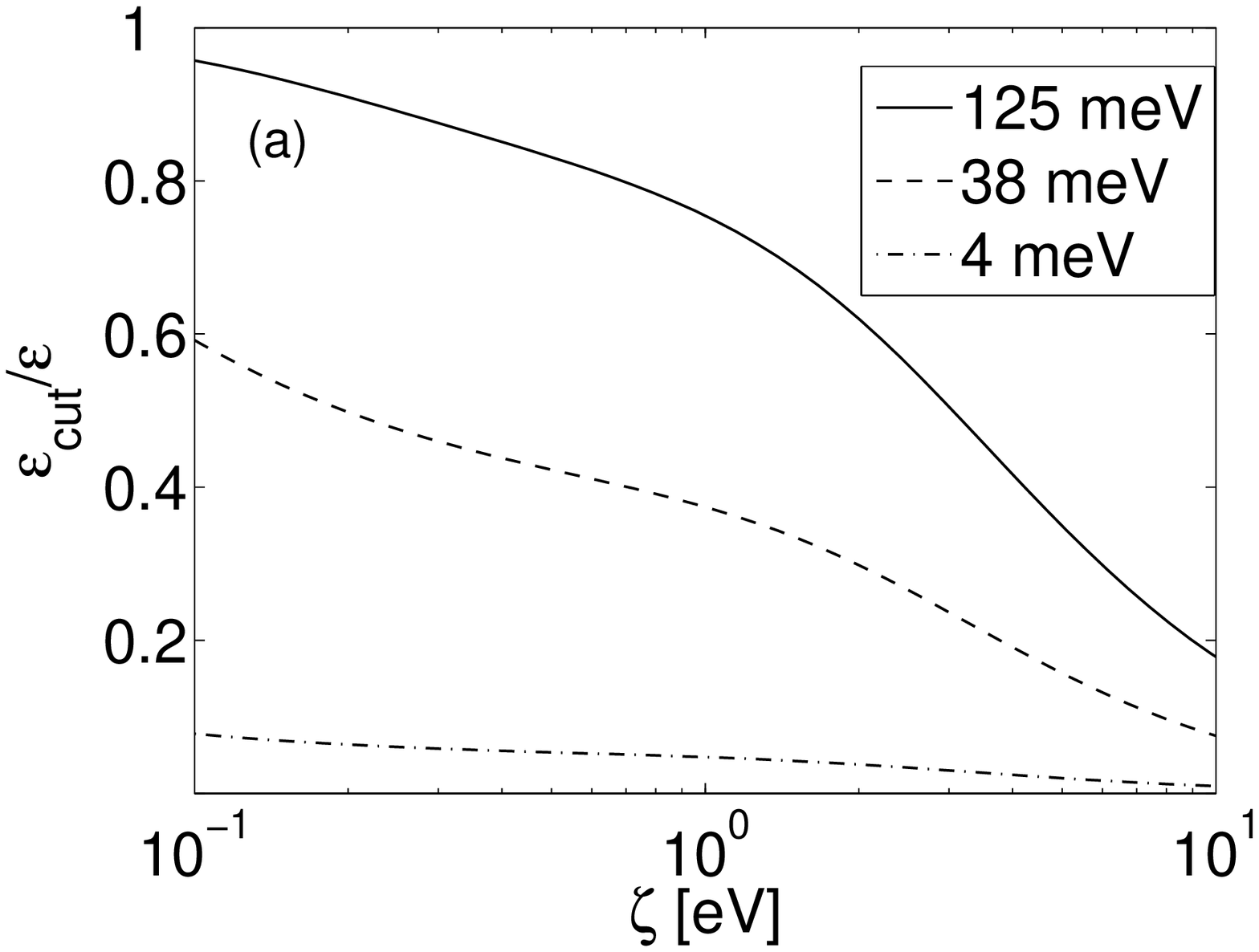}
\includegraphics[width=5.60cm]{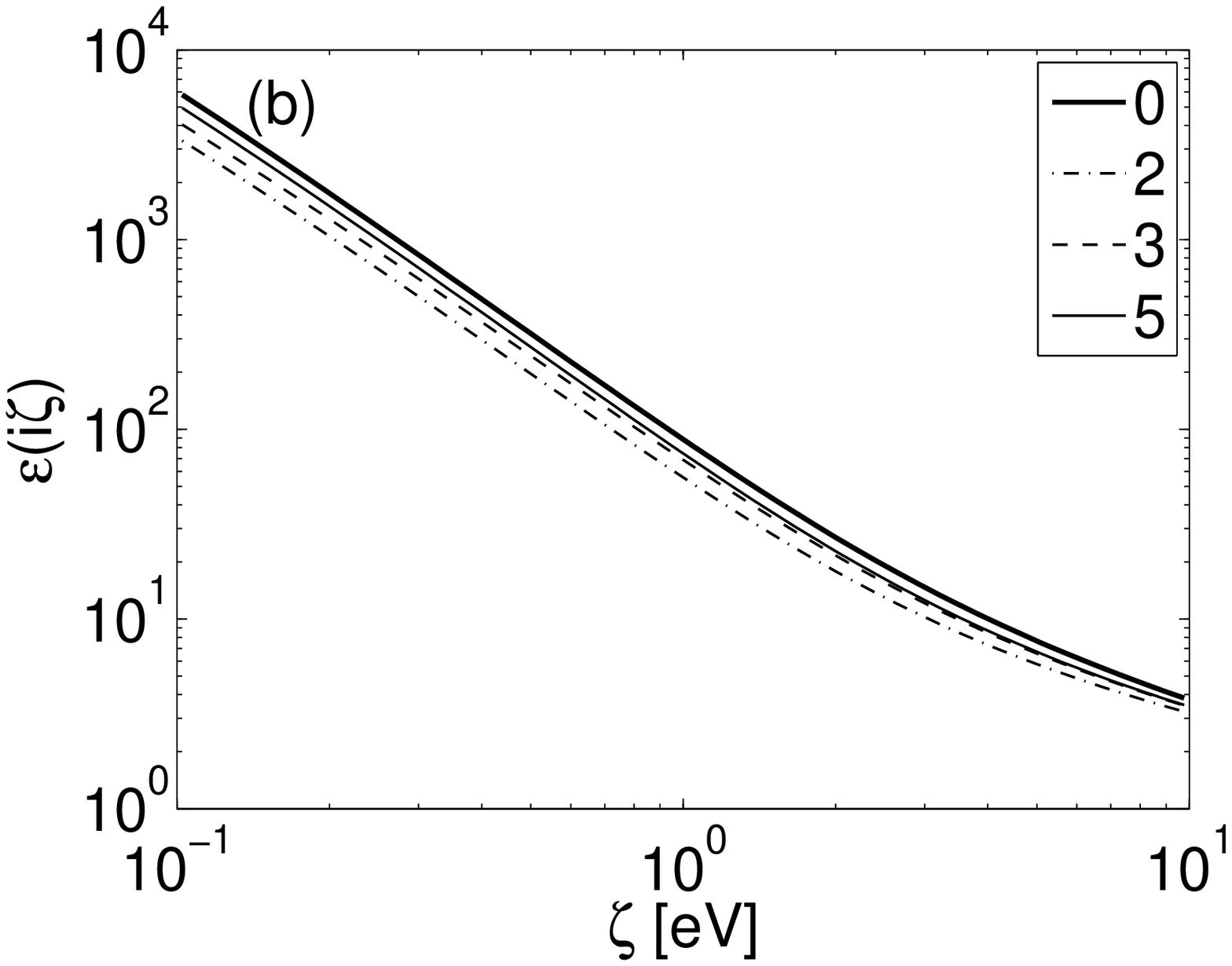}
\caption{(a) Relative contribution
$\varepsilon_{cut}(i\zeta)/\varepsilon(i\zeta)$ to the dielectric
function of gold at imaginary frequencies originating from the
extrapolated region $\omega<\omega_{cut}$ (see explanations in the
text). (b) The dielectric functions at imaginary frequencies for
samples 2,3, and 5; the thick curve marked as 0 corresponds to an
"ideal sample" with the plasma frequency of single crystal. }
\label{fig4}
\end{figure}

We can estimate now the contribution of $\varepsilon_{cut}(i\zeta)$
to $\varepsilon(i\zeta)$. For that we assume the Drude behavior at
$\omega<\omega_{cut}$ with the parameters $\omega_p=9.0
\;\textrm{eV}$ and $\gamma=35\;\textrm{meV}$ \cite{Lam00}. At higher
frequencies $\omega>\omega_{cut}$ we take the data from the handbook
\cite{HB1}, for which the cut-off frequency is
$\omega_{cut}=0.125\;\textrm{eV}$. These extrapolation and data were
used for interpretation most of the experiments, where the Casimir
force was measured. In Fig. \ref{fig4}(a) the solid curve is the
ratio $\varepsilon_{cut}(i\zeta)/\varepsilon(i\zeta)$ calculated
with these data. One can see that at $\zeta=1\;\textrm{eV}$ ($d\sim
100\;\textrm{nm}$) the contribution to $\varepsilon(i\zeta)$ from
the frequency range $\omega<\omega_{cut}$ is 75\%. It means, for
example, that if we change the Drude parameters, three fourths of
$\varepsilon(i\zeta)$ will be sensitive to this change and only one
forth will be defined by the measured optical data. Therefore, the
extrapolation procedure becomes very important for reliable
prediction of $\varepsilon(i\zeta)$.

The Drude parameters can vary from sample to sample due to different
density of defects. The plasma frequency is related with the density
of quasifree electrons $N$ as $\omega_p^2=4\pi Ne^2/m_e^{*}$, where
for gold $m_e^{*}\approx m_e$ is the effective mass of electron. The
value $\omega_p=9.0 \;\textrm{eV}$ is the maximal value of this
parameter, which corresponds to $N$ in a single crystal Au. In this
way $\omega_p$ was estimated in Ref. \cite{Lam00}. All deposited
films have smaller values of $\omega_p$ as one can see from Tab.
\ref{tab2} because the density of the films is smaller than that for
the single crystal material. Precise values of the Drude parameters
are extremely important for evaluation of $\varepsilon(i\zeta)$ and
finally for calculation of the force.

The dielectric function $\varepsilon(i\zeta)$ becomes less dependent
on the Drude parameters if the cut-off frequency is smaller. For
example, our optical data \cite{Sve08} were collected up to minimal
frequency $\omega_{cut}=38\;\textrm{meV}$ that is about four times
smaller than in the handbook data. The dashed curve in Fig.
\ref{fig4}(a) shows the ratio $\varepsilon_{cut}/\varepsilon$
calculated for our sample 3 with the Drude parameters
$\omega_p=7.84\;\textrm{eV}$ and $\gamma=49\;\textrm{meV}$. Now the
relative contribution of $\varepsilon_{cut}(i\zeta)$ at
$\zeta=1\;\textrm{eV}$ is 37\%. It is much smaller than for handbook
data, but still dependence on the precise Drude parameters is
important. Let us imaging now that we have been able to measure the
dielectric response of the material for the same sample 3 from
\cite{Sve08} to frequencies as low as $1\;\textrm{THz}$. In this
case the cut-off frequency is $\omega_{cut}=4\;\textrm{meV}$ and the
relative contribution of the extrapolated region
$\varepsilon_{cut}/\varepsilon$ is shown in Fig. \ref{fig4}(a) by
the dash-dotted line. Now this contribution is only 5\% at
$\zeta=1\;\textrm{eV}$.

The dielectric functions $\varepsilon_i(i\zeta)$, where $i=1,2,..,5$
is the number of the sample, were calculated using the Drude
parameters from Tab. \ref{tab2}. As a reference curve we use
$\varepsilon_0(i\zeta)$, which was evaluated with the parameters
$\omega_p=9.0 \;\textrm{eV}$ and $\gamma=35\;\textrm{meV}$ in the
range $\omega<0.125\;\textrm{eV}$ and at higher frequencies the
handbook data \cite{HB1} were used. The results are shown in Fig.
\ref{fig4}(b). As was expected the maximal dielectric function is
$\varepsilon_0(i\zeta)$, which corresponds in the Drude range to a
perfect single crystal. Even for the best sample 5 (annealed film on
mica) the dielectric function is 15\% smaller than
$\varepsilon_0(i\zeta)$ at $\zeta=1\;\textrm{eV}$. For samples 1 and
2 the deviations are as large as 40\%.

\subsubsection{The force between Au films}
\label{sec:2.2.3}

\begin{figure}[tbp]
\includegraphics[width=5.75cm]{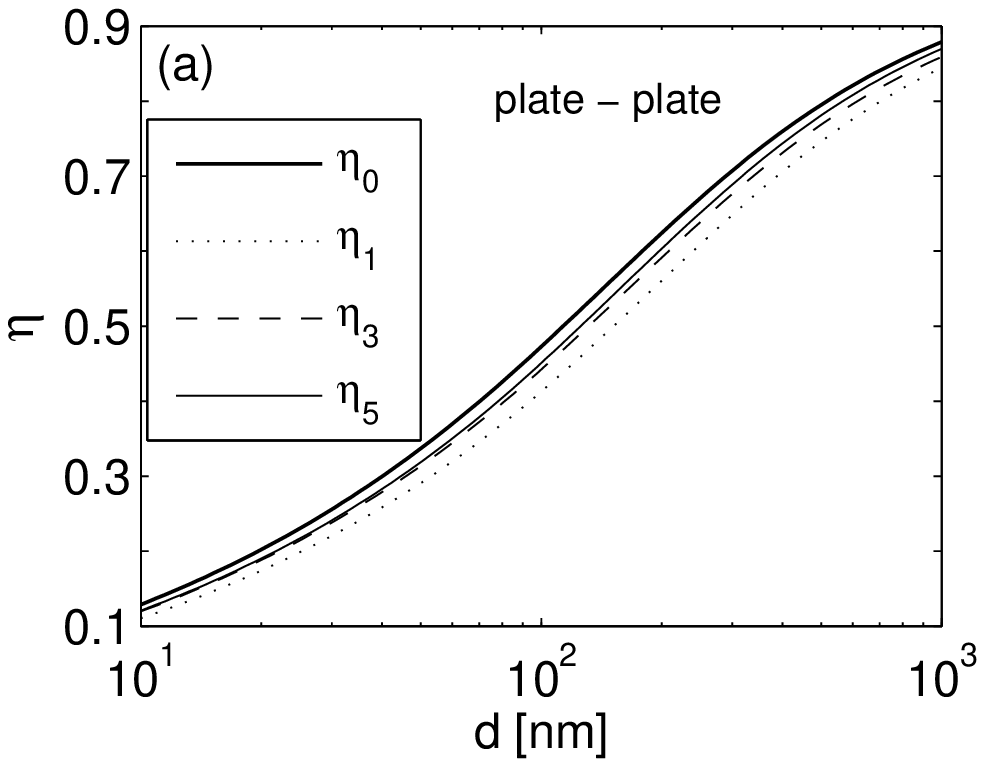}
\includegraphics[width=5.75cm]{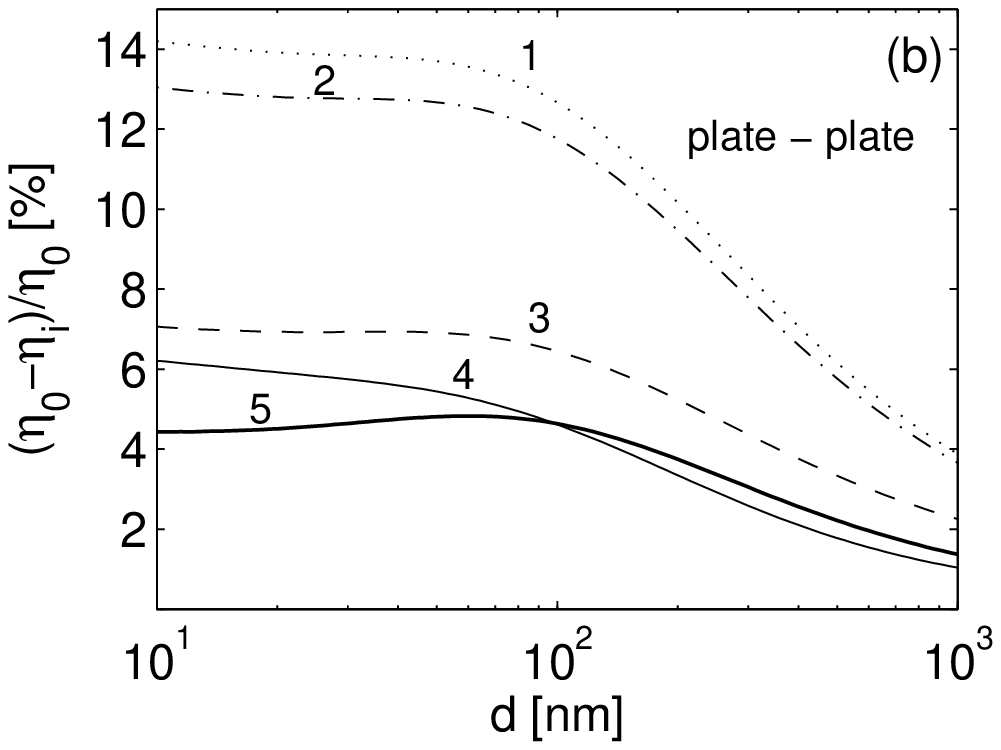}\newline
\caption{(a) Reduction factor $\eta$ as a function of the separation
$d$ for samples 1, 3, and 5. The thick line shows the reference
result calculated with $\varepsilon_{0}(i\zeta)$. (b) Relative
deviations of the reduction factors for different samples from the
reference curve $\eta_0(d)$, which were evaluated using the handbook
optical data \cite{HB1} and the Drude parameters
$\omega_p=9\;\textrm{eV}$, $\omega_{\tau}=35\;\textrm{meV}$. }
\label{fig5}
\end{figure}

It is convenient to calculate not the force itself but so called
the reduction factor $\eta$, which is defined as the ratio of the
force to the Casimir force between ideal metals:
\begin{equation}\label{eta_pp_def}
    \eta(d)=\frac{F(d)}{F^c(d)},\ \ \ F^c(d)=-\frac{\pi^2\hbar c}{240
    d^4}.
\end{equation}
We calculate the force between similar materials at $T=0$ using
the substitute (\ref{subs}) in Eq. (\ref{Lif_imag}). For
convenience of the numerical procedure one can make an appropriate
change of variables so that the reduction factor can be presented
in the form
\begin{equation}\label{eta_pp}
    \eta(d)=\frac{15}{2\pi^4}\sum\limits_{\mu=s,p}\int\limits_{0}^{1}dx
    \int\limits_{0}^{\infty}\frac{dyy^3}{r^{-2}_{\mu}e^y-1},
\end{equation}
where the reflection coefficients as functions of $x$ and $y$ are
defined as
\begin{equation}\label{refl_xy}
    r_s=\frac{1-s}{1+s},\ \ \
    r_p=\frac{\varepsilon(i\zeta_{c}xy)-s}{\varepsilon(i\zeta_{c}xy)+s},
\end{equation}
with
\begin{equation}\label{s_def}
    s=\sqrt{1+x^2\left[\varepsilon(i\zeta_{c}xy)-1\right]},\ \ \ \zeta_{c}=\frac{c}{2d}.
\end{equation}
The integral (\ref{eta_pp}) was calculated numerically with
different dielectric functions $\varepsilon_i(i\zeta)$. The results
are presented in Fig. \ref{fig5}(a) for samples 1, 3, and 5.  The
reference curve (thick line) calculated with $\varepsilon_0(i\zeta)$
is also shown for comparison. It represents the reduction factor,
which is typically used in the precise calculations of the Casimir
force between gold surfaces. One can see that there is significant
difference between this reference curve and those that correspond to
actual gold films. To see the magnitude of the deviations from the
reference curve, we plot in Fig. \ref{fig5}(b) the ratio
$(\eta_0-\eta_i)/\eta_0$ as a function of distance $d$ for all five
samples.

At small  distances the deviations are more sensitive to the value
of $\omega_p$. At large distances the sample dependence becomes
weaker and more sensitive to the value of $\omega_{\tau}$. For
samples 1 and 2, which correspond to the 400 nm and 200 nm films
deposited on Si substrates, the deviations are especially large.
They are 12-14\% at $d<100\;\textrm{nm}$ and stay considerable
even for the distances as large as 1 $\mu \textrm{m}$. Samples 3,
4, and 5 have smaller deviations from the reference case but even
for these samples the deviations are as large as 5-7\%.

\subsection{Low dielectric materials}
\label{sec:2.3}

\begin{figure}[tbp]
\begin{center}
\includegraphics[width=5.75cm]{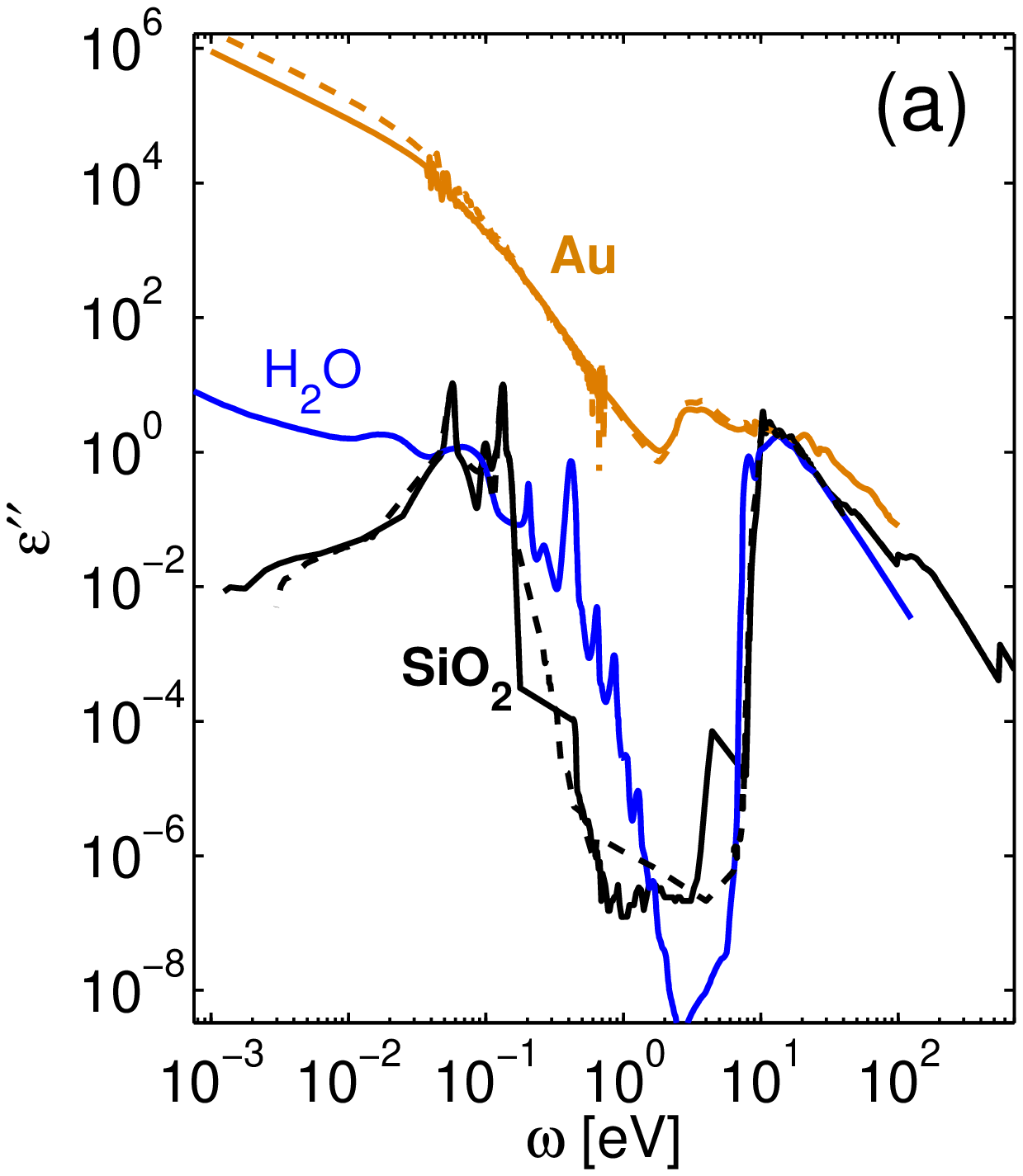}
\includegraphics[width=5.75cm]{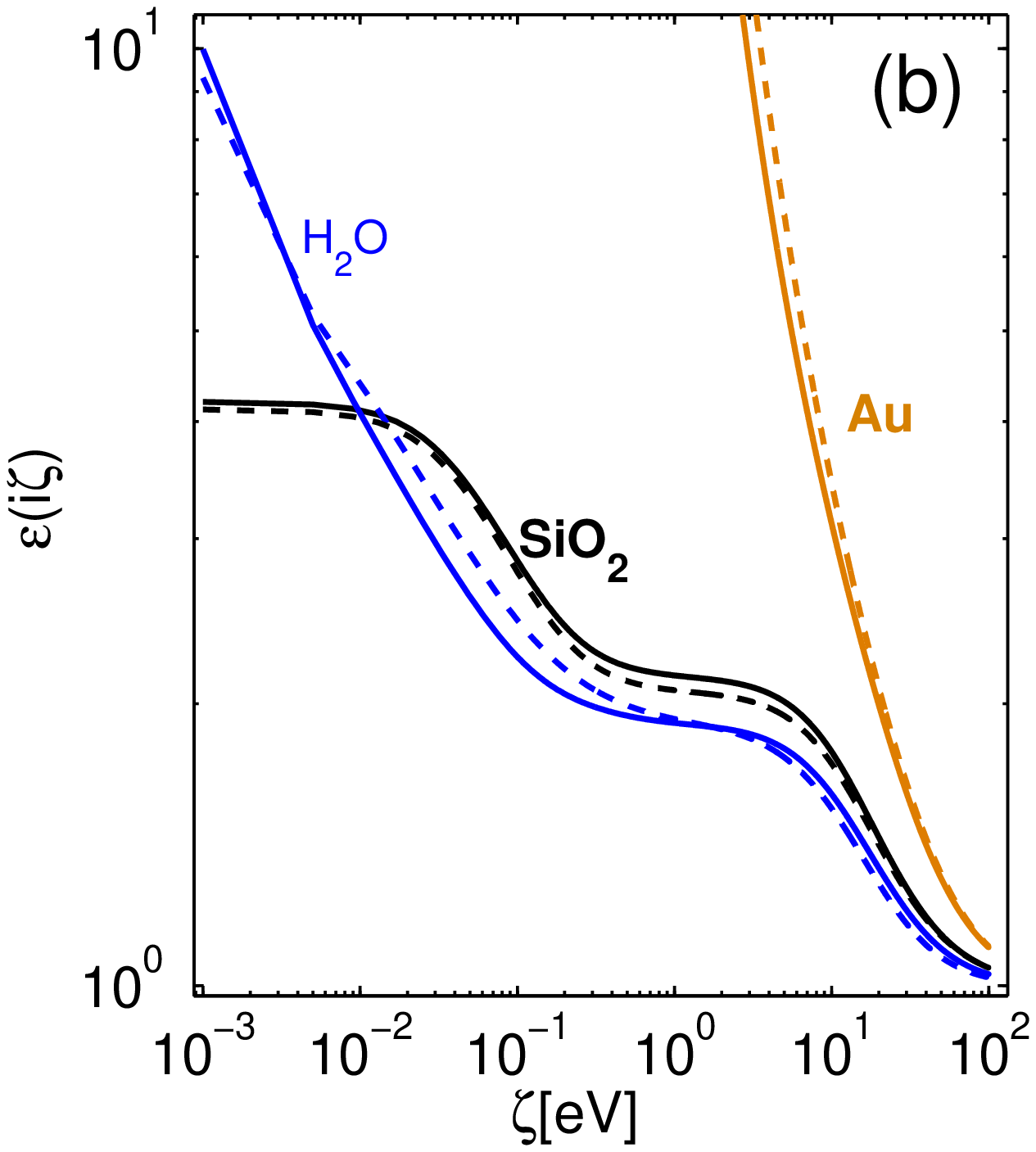}\newline
\caption{(a) Dielectric data of the materials obtained from
references in text. (b) Dielectric functions at imaginary
frequencies. The solid and dashed lines for silica and gold are two
different sets of optical data. For water the solid line is from the
data in Ref. \cite{Seg81}, and the dashed line is an 11-order
oscillator model \cite{Ngu00}, which has been fitted to a different
set of optical data. } \label{fig6}
\end{center}
\end{figure}

The Lifshitz theory predicts \cite{DLP} that the dispersive force
can be changed from attractive to repulsive by choosing the
interacting materials immersed in a liquid. Recently this prediction
was confirmed experimentally \cite{Mun09} (see Capasso paper in this
volume). Repulsive forces arise when the dielectric functions at
imaginary frequencies in the liquid gap, $\varepsilon_{0}(i\zeta)$,
is in between the functions of the interacting bodies 1 and 2:
$\varepsilon_{1}(i\zeta)>\varepsilon_{0}(i\zeta)>\varepsilon_{2}(i\zeta)$.
One can expect significant dependence on precise dielectric
functions nearby the transition from attractive to repulsive force.
This situation is exactly the case for the system
silica-liquid-gold. In this section we present calculations for
multiple liquids with various degrees of knowledge of the dielectric
functions.

Liquids do not have grains or defects, but the density of a liquid
is a function of temperature \cite{Dag00}, and as a result the
number of absorbers in the liquid varies with temperature.
Furthermore, liquids can contain impurities like salt ions which can
change the dielectric function (see discussion in Ref.
\cite{Mun08}). Although for metals the dielectric function is very
large in the IR regime, for liquids and glasses it is not the case.
Consequently for low dielectric materials the UV and VUV dielectric
data have a strongest effect on the forces.

For gold, silica, and water the dielectric functions are well known
and measured by various groups. Let us consider first the
interaction in the system gold-water-silica. We will use two sets of
data for gold from the previous subsection (sample 1 and the "ideal
sample"). Also two sets of data will be used for silica as obtained
by different groups \cite{Kit07}. Finally, for water we will use the
data of Segelstein compiled from different sources \cite{Seg81}, and
an 11-order oscillator model \cite{Ngu00} that has been fit to
different sets of data \cite{Par81,Rot96}. All the dielectric data
are collected in Fig. \ref{fig6}(a). The corresponding functions at
imaginary frequencies are shown in Fig. \ref{fig6}(b). One can see
considerable difference between solid and dashed curves
corresponding to different sets of the data.

\begin{figure}[tbp]
\includegraphics[width=5.75cm]{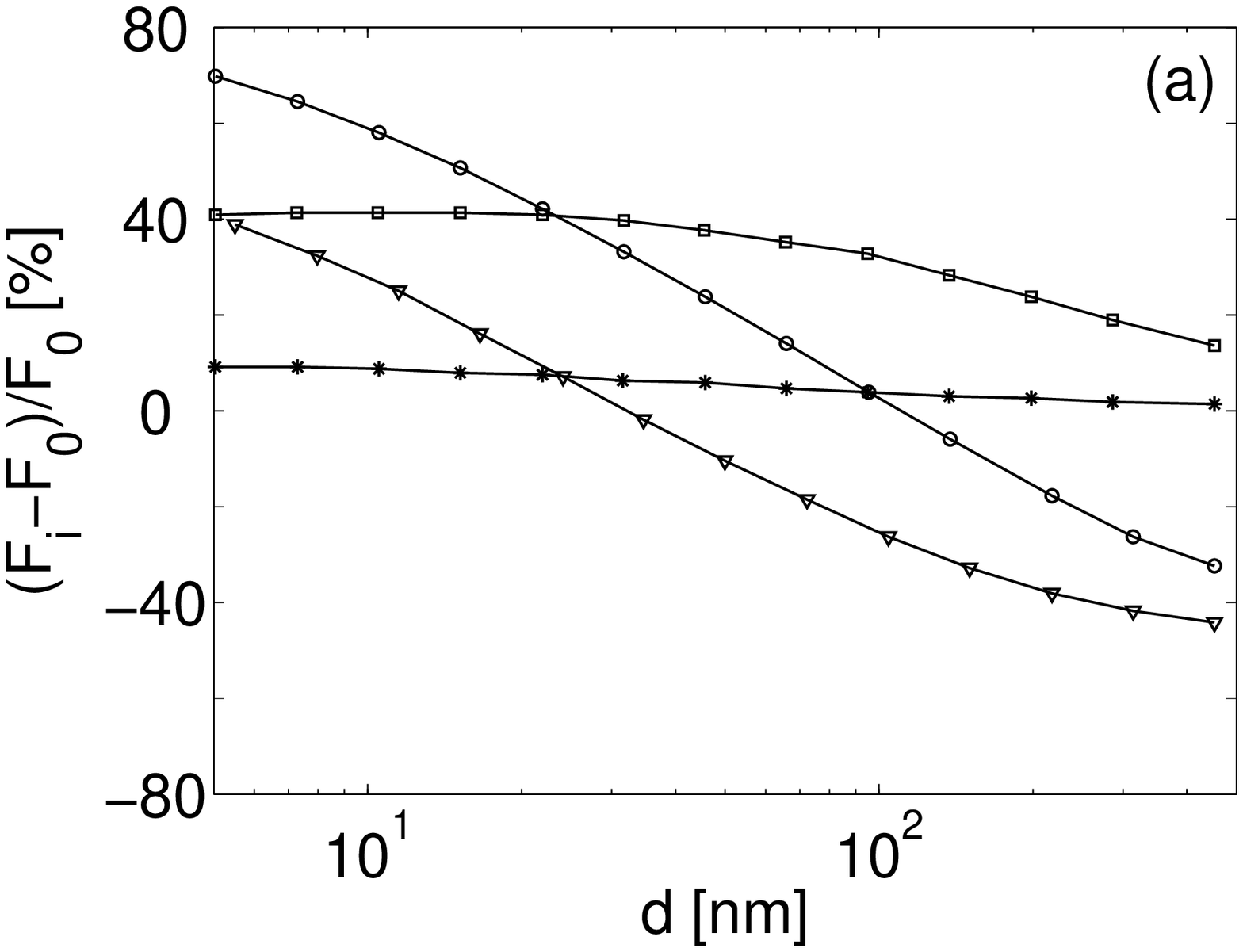}
\includegraphics[width=5.75cm]{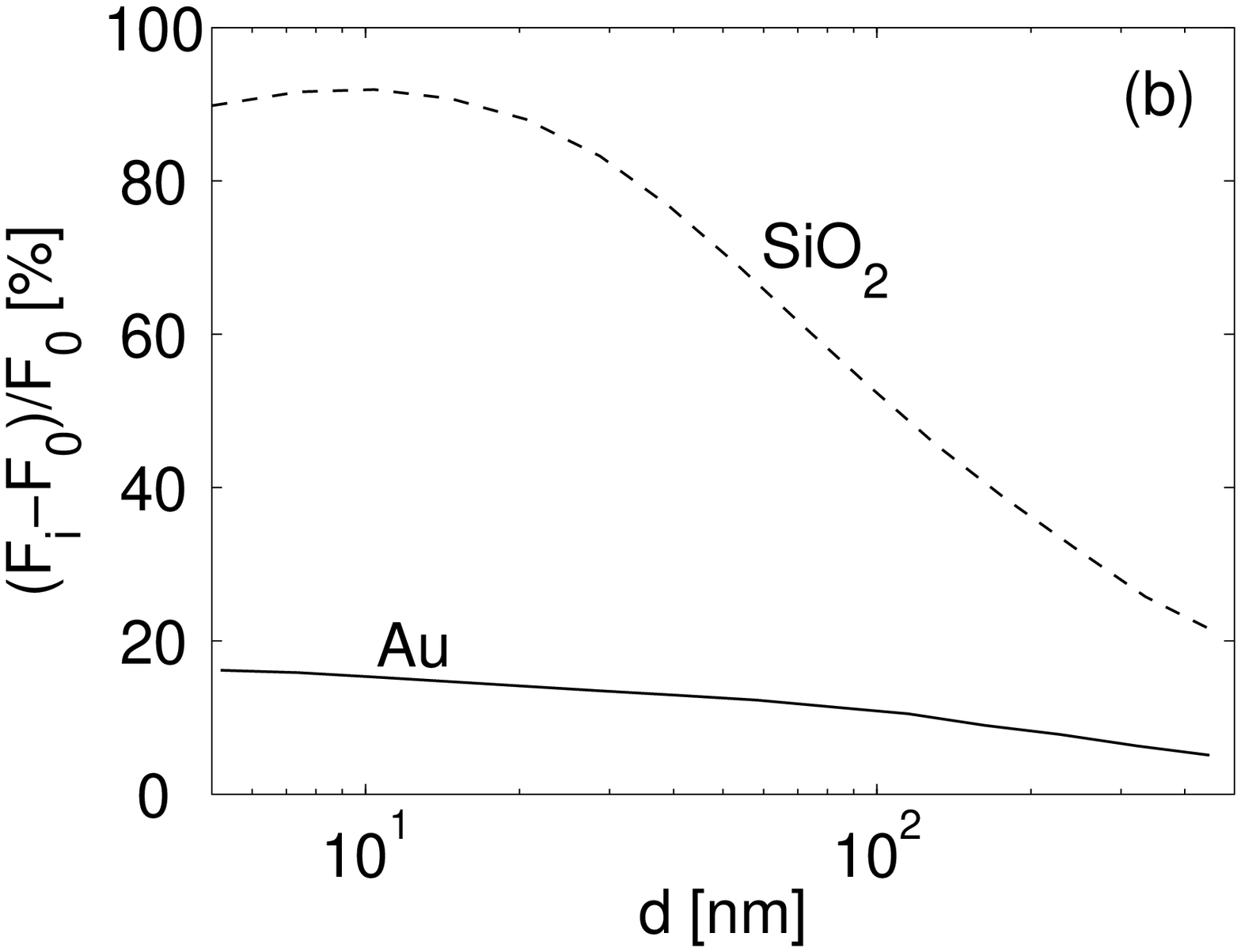}\newline
\caption{(a) Variation of the force relative to $F_0(d)$ in the
system gold-water-silica. Circles, squares, triangles and stars mark
the curves which were calculated using different sets of the
dielectric data. (b) Variation of the force for silica-water-silica
(dashed line) with different sets of data for SiO$_2$ and for
gold-water-gold (solid line) using different data for Au.}
\label{fig7}
\end{figure}

It has to be noted that $\varepsilon^{''}(\omega)$ for water and
silica are very close in a wide range of frequencies $5\cdot
10^{-2}<\omega<5\cdot 10^{2}\;\textrm{eV}$. As the result at
imaginary frequencies $\varepsilon_{SiO_{2}}(i\zeta)$ and
$\varepsilon_{H_{2}O}(i\zeta)$ differ on 30\% or less in the range
$10^{-2}<\zeta<10^{2}\;\textrm{eV}$, which is comparable with the
magnitude of variation of $\varepsilon(i\zeta)$ due to data
scattering. This similarity in the dielectric functions results in a
strong dependence of the Casimir force in the system
gold-water-silica on the used optical data. It is illustrated in
Fig. \ref{fig7}(a) where the relative change of the force is shown.
The scatter of the force riches of the level of 60\% for separation
$d<500\;\textrm{nm}$. Even more clear the effect can be seen in Fig.
\ref{fig7}(b). The solid curve shows variation of the force in
Au-water-Au system when different optical data for Au are used. In
this case the relative change of the force is not very large.
However, for the system silica-water-silica the use of different
optical data for silica influence the force very significantly
(dashed curve).

We have to conclude that comparison of force measurements with
prediction of the Lifshitz theory becomes reliable when the
dielectric properties of the specific samples used in force
measurement are measured over a wide range of frequencies.

In most of the papers where liquid gap between bodies is studied
the dielectric function of the liquid is approximated using the
oscillator models \cite{Oss88,Mil96}. For illustration purposes we
mention that alcohols (and other liquid substances) can be
described, for example, by a three oscillator model for the
dielectric function $\varepsilon(i\zeta)$ at imaginary frequencies
\cite{Oss88}
\begin{equation}\label{3osc}
    \varepsilon(i\zeta)=1+\frac{\varepsilon_0-\varepsilon_{IR}}
    {1+\zeta/\omega_{MW}}+\frac{\varepsilon_{IR}-n_0^2}{1+
    (\zeta/\omega_{IR})^2}+\frac{n_0^2-1}{1+
    (\zeta/\omega_{UV})^2}.
\end{equation}
Here $n_0$ is the refractive index in the visible range,
$\varepsilon_0$ is the static dielectric constant, and
$\varepsilon_{IR}$ is the dielectric constant where MW relaxation
ends and IR begins. The parameters $\omega_{MW}$, $\omega_{IR}$, and
$\omega_{UV}$ are the characteristic frequencies of MW, IR, and UV
absorption, respectively. It has to be stressed that oscillator
models should be used with caution, because some of them are of poor
quality \cite{Zwo09a}.

\begin{figure}[tbp]
\begin{center}
\includegraphics[width=11.5cm]{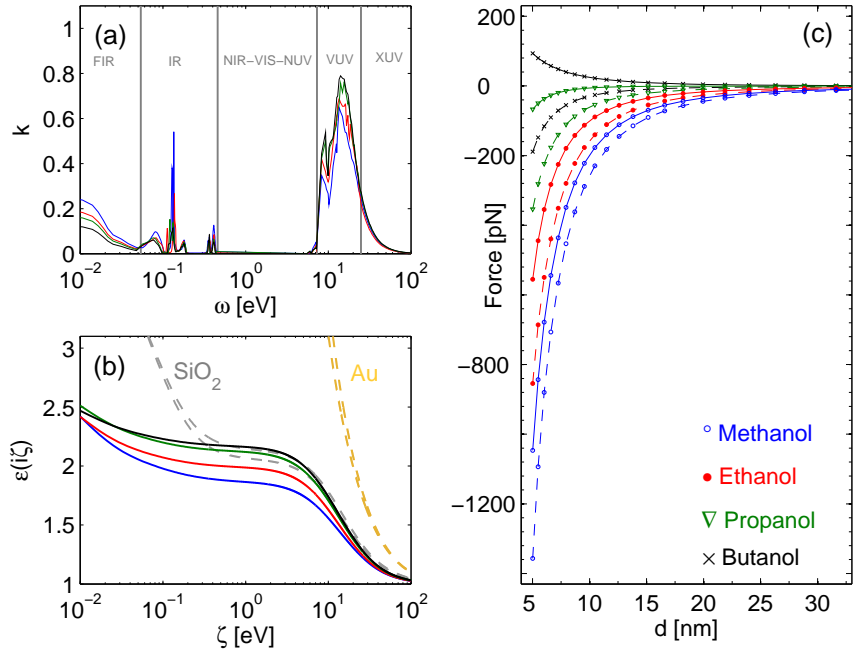}\newline
\caption{Dielectric data at real (a) and at imaginary frequencies
(b) for methanol, ethanol, propanol, and butanol, as described in
the text. For comparison the dielectric data for silica and gold
samples are also shown in (b). In (c) we show the forces for the
gold-alcohol-silica, for two different sets of measured dielectric
data for silica. In the case of butanol the force is attractive for
one set, and repulsive for another one.} \label{fig8}
\end{center}
\end{figure}

For ethanol rather detailed information on the dielectric function
exists, but even in this case variation in dielectric data was found
\cite{Zwo09a,Fen02}. An interesting fact for higher alcohols is that
the absorption in the UV range increases when increasing the alkane
chain. In Fig. \ref{fig8} we show the dielectric data for the first
four alcohols.

The VUV data were taken from \cite{Oga58}. These measurements were
done in the gas phase, but they can be converted to the liquid case
by considering the number of absorbers in gas and liquid. The near
UV data was taken from \cite{Sal71}. For the XUV we have only data
for ethanol and propanol \cite{Koi86}. For methanol and butanol we
used cubic extrapolation, $\varepsilon^{''}\sim 1/\omega^3$, which
is in very good agreement with the cases of ethanol and propanol. In
the near IR (NIR) to visible (VIS) ranges the extinction coefficient
$k$ of ethanol, and other alcohols, is very low and can be taken to
be zero, $k=0$, which is qualitatively consistent with the fact that
all alcohols are transparent in the visible range. The IR data can
be found in Ref. \cite{Set79}. The far IR (FIR) data are known only
for methanol \cite{Ber95}. For the other alcohols we take the
similar functional behavior as for methanol but with different
parameters and extrapolate the data to far IR in this way.

If one has to estimate the dielectric functions for some alcohols,
first of all one has to have measured data in the range of major
IR peaks and even more importantly the UV absorption must be
carefully measured. Thus the dielectric functions at imaginary
frequencies should be reasonably accurate to within the scatter of
the data as it was found for ethanol \cite{Fen02}.

With the optical data for alcohols we calculated the forces in the
system gold-alcohol-silica. The forces are attractive and become
weaker for methanol, ethanol and propanol. For butanol they are
extremely weak, but still either repulsive or attractive. Caution is
required in the analysis of optical properties in liquids since in
general the KK consistency has to be applied properly in order to
correct for variation of the dielectric properties observed in
between different measuring setups. Effectively the force for
gold-butanol-silica is screened to within the scatter of the forces
related to sample dependence of the optical properties of silica.
Measurements between gold and glass with simple alcohols were
performed, but experimental uncertainty, and double layer forces
prevented the measurement of this effect \cite{Zwo09b}.

\section{Influence of surface roughness on the Casimir-Lifshitz force}
\label{sec:3}

The Lifshitz formula (\ref{Lif_imag}) does not take into account
inevitable roughness of the interacting bodies. When rms roughness
of the bodies is much smaller than the separation, then the
roughness influence on the force can be calculated using the
perturbation theory \cite{Gen03,Mai05a,Mai05b}. However, when the
separation becomes comparable with the roughness the perturbation
theory cannot be applied. It was demonstrated experimentally
\cite{Zwo08b} that in this regime the force deviates significantly
from any theoretical prediction. The problem of short separations
between rough bodies is one of the unresolved problems. In this
section we give introduction into interaction of two rough plates
and a sphere and a plate.

\subsection{Main characteristics of a rough surface}
\label{sec:3.1}

Suppose there is a rough plate which surface profile can be
described by the function $h(x,y)$, where $x$ and $y$ are the
in-plane coordinates. An approximation for this function provides,
for example, an AFM scan of the surface. It gives the height
$h_{ij}$ at the pixel position $x_i=\Delta\cdot i$ and
$y_j=\Delta\cdot j$, where $i,j=1,2,...,N$ and $\Delta$ is the pixel
size related with the scan size as $L=\Delta\cdot N$. We can define
the mean plane of the rough plate as the averaged value of the
function $h(x,y)$: $\bar{h}=A^{-1}\int dxdy h(x,y)$, where $A$ is
the area of the plate. This definition assumes that the plate is
infinite. In reality we have to deal with a scan of finite size, for
which the mean plane is at
\begin{equation}\label{aver}
    h_{av}=\frac{1}{N^2}\sum\limits_{i,j} h(x_i,y_j).
\end{equation}
The difference $\bar{h}-h_{av}$, although small, is not zero and is
a random function of the scan position on the plate. This difference
becomes larger the smaller the scan size is. Keeping in mind this
point, which can be important in some situations (see below), we can
consider (\ref{aver}) as an approximate definition of the mean plane
position.

An important characteristic of the a rough surface is the rms
roughness $w$, which is given as
\begin{equation}\label{rms_rough}
    w=\frac{1}{N^2}\sum\limits_{i,j} \left[h(x_i,y_j)-h_{av}\right]^2.
\end{equation}
It has the meaning of the surface width. More detailed information
on the rough surface can be extracted from the height-difference
correlation function defined for the infinite surface as
\begin{equation}\label{corr_fun}
    g(R)=\frac{1}{A}\int dxdy\;
    \left[h(\textbf{r}+\textbf{R}) - h(\textbf{r})\right]^2,
\end{equation}
where $\textbf{r}=(x,y)$ and $\textbf{R}=\textbf{r}'-\textbf{r}$.

A wide variety of surfaces, as for example, deposited thin films far
from equilibrium, exhibit the so called self-affine roughness which
is characterized besides the rms roughness amplitude $w$ by the
lateral correlation length $\xi$   (indicating the average lateral
feature size), and the roughness exponent $0<H<1$
\cite{Mea91,Kri95,Zha01}. Small values of $H\sim 0$ corresponds to
jagged surfaces, while large values of $H\sim 1$ to a smooth hill
valley morphology. For the self-affine rough surfaces $g(R)$ scales
as
\begin{equation}\label{asym}
    g(R)=\left\{
                  \begin{array}{c}
                    R^{2H},\ R\ll \xi, \\
                    2w^2,\ R\gg \xi. \\
                  \end{array}
    \right.
\end{equation}
The parameters $w$, $\xi$ and $H$ can be determined from the
measured height-difference correlation function $g(R)$. This
function can be extracted approximately from the AFM scans of the
surface.

To find roughness correction to the force one has to know (see
below) the spectral density $\sigma(k)$ of the height-height
correlation function $C(R)$. The latter is related with $g(R)$ as
$g(R)=2w^2-C(R)$. An analytic form of the spectral density for a
self-affine surface is given by \cite{Pal93}
\begin{equation}\label{spectr_dens}
    \sigma(k)=\frac{CHw^2\xi^2}{\left(1+k^2\xi^2\right)^{1+H}}, \ \
    \ C=\frac{2}{1-\left(1+k_c^2\xi^2\right)^{-H}}.
\end{equation}
Here $C$ is a normalization constant \cite{Pal93, Zha01} and
$k_c=2\pi/L_c$ is the cutoff wavenumber.

\subsection{Roughness correction}
\label{sec:3.2}

While the separation between two surfaces is large in comparison
with the rms roughness, $d\gg w$, one can use the perturbation
approach to calculate the roughness correction to the Casimir force.
This correction was calculated first using the proximity force
approximation \cite{Kli96}. This approximation assumes that the
surface profile slowly varied in comparison with the distance
between the bodies. The lateral size of a rough profile is given by
the correlation length $\xi$, therefore, PFA can be applied if
$\xi\gg d$. This condition is very restrictive since typical values
of $\xi$ for deposited metals films (grain size) are in the range
20-100 nm. In most of the experimental situations the condition
$\xi\gg d$ is broken. It was realized for the first time in Ref.
\cite{Gen03}. More general theory \cite{Gen03,Mai05a,Mai05b} for the
roughness correction can be applied at $\xi \leq d$ and treats the
correction perturbatively within the scattering formalism (see
Lambrecht paper in this volume). Here we discuss application of this
theory to realistic rough surfaces and describe situations, for
which one has to go beyond the perturbation theory to find agreement
with experiments.

\subsubsection{Application of the perturbation theory}
\label{sec:3.2.1}

\begin{figure}[tbp]
\begin{center}
\includegraphics[width=9cm]{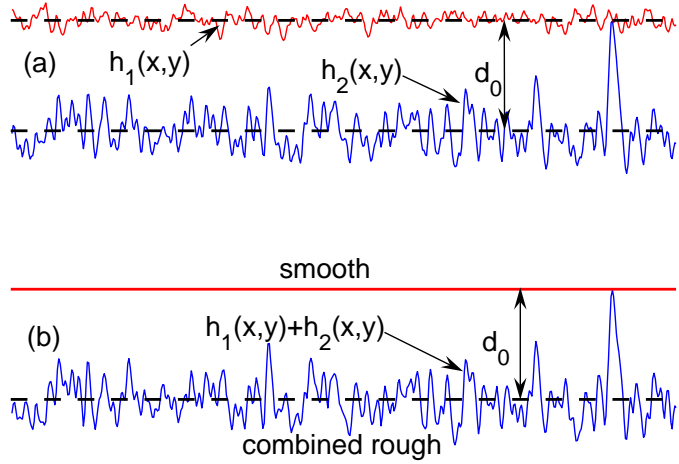}\newline
\caption{Contact of two rough surfaces. (a) Two rough plates in
contact. Roughness of each plate, $h_i(x,y)$, is counted from the
mean plane shown by the dashed line. The distance between bodies is
the distance between these mean planes. (b) The interaction between
two rough plates is equivalent to the interaction between a smooth
plate and a rough plate with the roughness given by the combined
profile $h(x,y)$. The distance upon contact, $d_0$ has well defined
meaning in this case. See Sec. \ref{sec:3.3} and Ref. \cite{Zwo09c}
for details.} \label{fig9}
\end{center}
\end{figure}

Let us consider two parallel rough plates. A plate surface can be
described by the roughness profile $h_i(x,y)$ ($i=1,2$ for plate 1
or 2) as shown in Fig. \ref{fig9}(a). The averaged value over large
area is assumed to be zero $\left\langle h_i(x,y)\right\rangle=0$.
Then the local distance between the plates is
\begin{equation}\label{loc_dist}
    d(x,y)=d-h_1(x,y)-h_2(x,y).
\end{equation}
This distance depends on the combined rough profile
$h(x,y)=h_1(x,y)+h_2(x,y)$. As explained in Sec. \ref{sec:3.3} the
interaction of two rough plates is equivalent to the interaction of
a smooth plate and a rough plate with the roughness given by the
combined profile $h(x,y)$ (see Fig. \ref{fig9}(b)).

Let us assume further that the interaction energy per unit area of
two flat plates is $E_{pp}(d)$. If the rms roughness of the combined
profile $h(x,y)$ is small, $w\ll d$, but the correlation length is
large, $\xi\gg d$, we can present the interaction between rough
plates as
\begin{equation}\label{rough_inter}
    E_{pp}^{rough}=\left\langle E_{pp}\left(d(x,y)\right)\right\rangle
    \approx E_{pp}(d)+\frac{E_{pp}^{\prime\prime}}{2}\left\langle
    h^2\right\rangle,
\end{equation}
where $\left\langle h^2\right\rangle=w^2=w_1^2+w_2^2$. Equation
(\ref{rough_inter}) defines the PFA roughness correction $\delta
E_{pp}=E_{pp}^{\prime\prime}w^2/2$. This correction was used in all
early studies to estimate the roughness effect.

It was noted \cite{Gen03} that in most experimental configurations
the condition $\xi\gg d$ is broken and PFA cannot be applied. In
Refs. \cite{Mai05a,Mai05b} was developed a theory, which is not
restricted by the condition $\xi\gg d$. Within this theory the
roughness correction is expressed via the spectral density of the
rough surface $\sigma(k)$ as
\begin{equation}\label{rough_inter}
    \delta E_{pp}(d)=\int\frac{d^2k}{(2\pi)^2}G(k,d)\sigma(k),
\end{equation}
where $G(k,d)$ is a roughness response function derived in
\cite{Mai05b}. The PFA result (\ref{rough_inter}) is recovered from
here in the limit of small wavenumbers $k\rightarrow 0$ when
$G(k,d)\rightarrow E_{pp}^{\prime\prime}(d)/2$. The roughness power
spectrum is normalized by the condition $\int
d^2k\sigma(k)/(2\pi)^2=w^2$. The spectrum itself can be obtained
from AFM scans and in the case of self-affine rough surfaces is
given by Eq. (\ref{spectr_dens}).

Let us enumerate the conditions at which Eq. (\ref{rough_inter}) is
valid. (i) The lateral dimensions of the roughness $\xi $ must be
much smaller than the system size $L$, $\xi\ll L$. This is usually
the case in experiments. (ii) The rms roughness $w$ must be small
compared to the separation distance, $w\ll d$. This condition means
that roughness is treated as perturbative effect. (iii) The lateral
roughness dimensions must be much larger than the vertical
dimensions, $w\ll\xi$ \cite{Mai05b}. The last two assumptions are
not always satisfied in the experiment.

In the plate-plate configuration the force per unit area can be
calculated as the derivative of $E_{pp}^{rough}(d)$. For the
sphere-plate configuration, which is used in most of the
experiments, the force is calculated with the help of PFA as
$F_{sp}(d)=2\pi R E_{pp}^{rough}(d)$. In contrast with the roughness
correction the latter relation is justified for $d\ll R$, which
holds true for most of the experimental configurations. We use the
sphere-plate configuration to illustrate the roughness effect. The
deposited gold films can be considered as self-affine. For all
calculations reported here we are using our smoothest spheres with
the parameters $w=1.8$ nm, $\xi=22$ nm, and $H=0.9$. We alter only
the plate roughness since it is easy to prepare and replace during
experiments. We use the optical data for gold films described
previously. It was found that the PFA limit is quickly recovered for
increasing correlation length. Deviations from PFA prediction for
real films were found to be about 1-5\% in the range $d=50-200$ nm.

Therefore, for real rough surfaces the scattering theory gives a few
percent correction to the force compared to the PFA. This difference
is difficult to measure. However, at small separations both PFA and
perturbation theory fail since the rms roughness becomes comparable
in size to the separation distance. It would be interesting to
calculate the roughness effect when $d$ is comparable with $w$. At
the moment there is no a theoretical approach to estimate the effect
except a direct numerical analysis similar to that used in
\cite{Rod07}. It will therefore be interesting to do a full
numerical analysis for films with high local slopes instead of using
perturbation theory. On the other hand it is experimentally possible
to go to sufficiently small distances as it will be discussed below.

\subsubsection{Experimental evidence of large roughness effect}
\label{sec:3.2.2}

The Casimir forces between a 100 $\mu$m gold coated sphere and
substrates covered with Au to different thicknesses from 100 nm to
1600 nm were measured in \cite{Zwo08b}. Different layers of Au
resulted to different roughnesses and different correlation lengths,
which are collected in Tab. \ref{tab3}. The roughness exponent was
constant $H=0.9\pm 0.05$ in agreement with former growth studies of
thin films \cite{Pal93,Kri95}. The sphere was attached to a
cantilever with a spring constant of 0.2 N/m. The calibration
procedure is described in \cite{Zwo08b}.

\begin{figure}[tbp]
\begin{center}
\includegraphics[width=9cm]{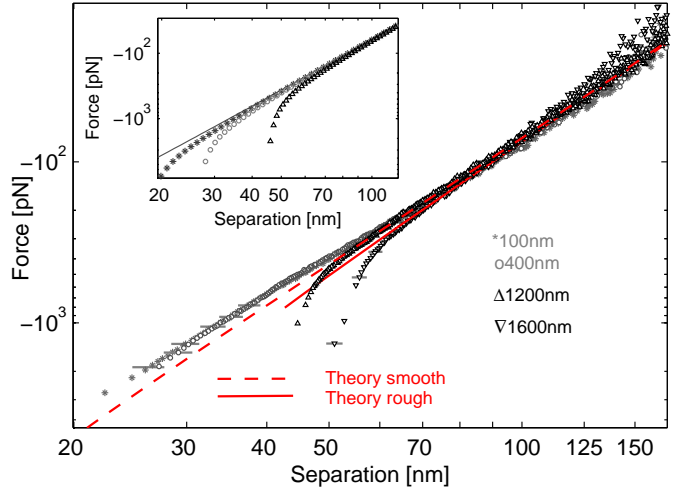}\newline
\caption{Casimir forces measured for the films of different
roughness. The roughness effect manifests itself as a strong change
in scaling at smaller separations, where the forces become much
stronger. Errors in separation are shown for some points by the
horizontal bars. The theoretical curves are shown for the 100 nm
(smooth) and 1600 nm (rough) films. The inset shows the forces
calculated by integrating over the roughness scans using PFA (see
text).} \label{fig10}
\end{center}
\end{figure}

The force results are shown in Fig. \ref{fig10}. Our measurements
were restricted to separations below 200 nm where the Casimir force
is large enough compared to the approximately linear signal from
laser light surface backscattering \cite{Har00,Zwo08b}. The small
separation limit or contact point is restricted by the jump to a
contact ($\sim 5$ nm) \cite{Zwo08c,Del07,Zwo07} and surface
roughness. Note that an error of 1.0 nm in absolute distance leads
to errors in the forces of up to 20\% at close separations as for
example $d\sim 10$ nm \cite{Zwo08b}. Thus we cannot detect the
scattering effects described above. What we do see is the failure of
the perturbation theory, for the roughest films, for which the
roughness strongly increases the force. These deviations are quite
large, resulting in much stronger forces at the small separations
$<70$ nm. At larger separations, 70-130 nm (within our measurement
range), where the roughness influence is negligible,  the usual
$1/d^{2.5}$ scaling of the force observed also in other experiments
with gold is recovered and agreement with the theory is restored.
For the smoother films deviations from theory below 40 nm can be
explained with the error in the distance.

\begin{figure}[tbp]
\begin{center}
\includegraphics[width=9cm]{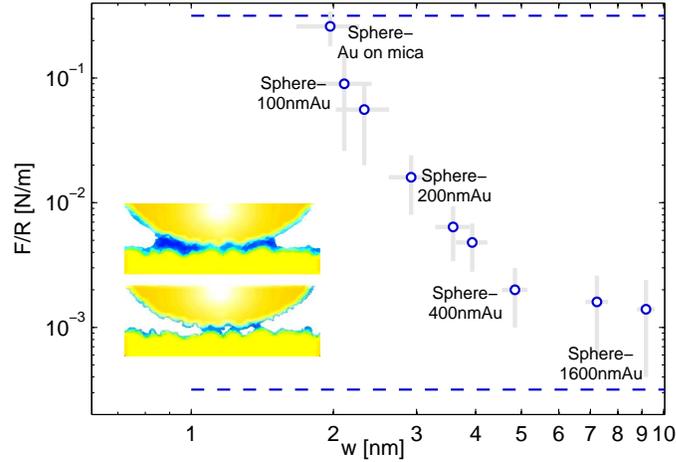}\newline
\caption{Capillary forces in air (relative humidity 2-60\%) for a
smooth gold coated sphere, $w_{sph}\sim 1$ nm, measured with a stiff
cantilever, $k=4$ N/m, and different rough films. The inset shows a
fully wetted sphere (upper dashed line), and a roughness asperity
wetted sphere (lower dashed line).} \label{fig11}
\end{center}
\end{figure}

Qualitatively the roughness effect can be reproduced by calculating
the force between small areas on the surfaces separated by the local
distance $d(x,y)$. One can call this procedure the non-perturbative
PFA approach. Although it is qualitative, it can be used to obtain
an estimate of the force at close proximity (2 nm above the point
upon contact), where the roughness has an enormous influence on the
Casimir force (see inset in Fig. \ref{fig10}). This explains the
jump to contact only partially, since approximately 5 nm above the
point of contact, the capillary force will act as well. This force
appears due to absorbed water and Kelvin condensation
\cite{Isr92,Mab97} resulting in water bridges formation between
bodies. In the limit of fully wetted surface (see Fig. \ref{fig11})
the capillary force is given by $F_{cap}\approx 4\pi\gamma
R\cos\vartheta$  (upper dashed line), while for a single asperity
(of size $\xi$ ) wetting the minimum capillary force is
$F_{cap}\approx 4\pi\gamma\xi \cos\vartheta$ (lower dashed line).
Here $\gamma$ is the surface tension of liquid, and $\vartheta$ is
the contact angle \cite{Zwo07,Zwo08c}.

At this point we have to compare the Casimir adhesion between rough
films with adhesion by capillary forces \cite{Del07,Zwo07}. While
Casimir forces may lead to stiction between movable parts, once the
surfaces are in contact capillary forces (being present in air
between hydrophilic surfaces into close proximity) are much
stronger. The roughness effect on capillary adhesion is also much
stronger as shown in Fig. \ref{fig11}. Note that the Casimir force
for a $R=50\;\mu$m sphere is in the order of 10 nN at 10 nm
separations. Capillary forces between a mica substrate and the same
sphere are as large as $10\;\mu$N deeming contact measurements with
soft cantilevers (spring constant $<1$ N/m) even impossible since
the retraction range is outside of that of most piezo z-ranges.
Measurements of the capillary forces with a smooth sphere used for
the Casimir force measurement are shown in Fig. \ref{fig11}. One can
see that when roughness increases a few times the force decreases by
more than a factor of 100. This can be related to full sphere
wetting and asperity wetting. The size of the sphere $R=50\;\mu$m is
1000 times larger than that of an asperity $\xi\sim50$ nm. Multiple
asperity capillary bridge formation is likely to happen in the rough
regime giving increasing forces.

\begin{figure}[tbp]
\begin{center}
\includegraphics[width=5.6cm]{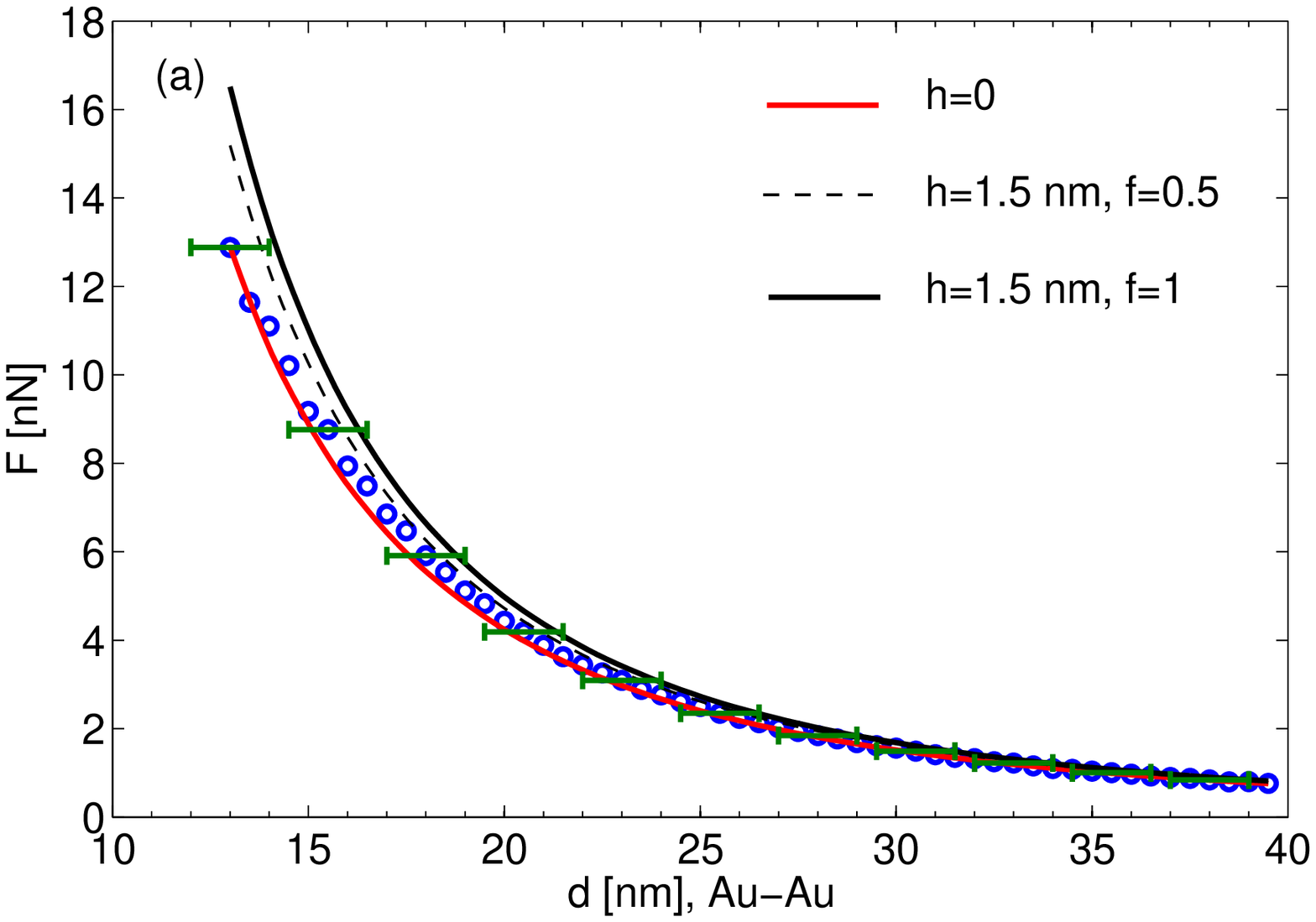}
\includegraphics[width=5.75cm]{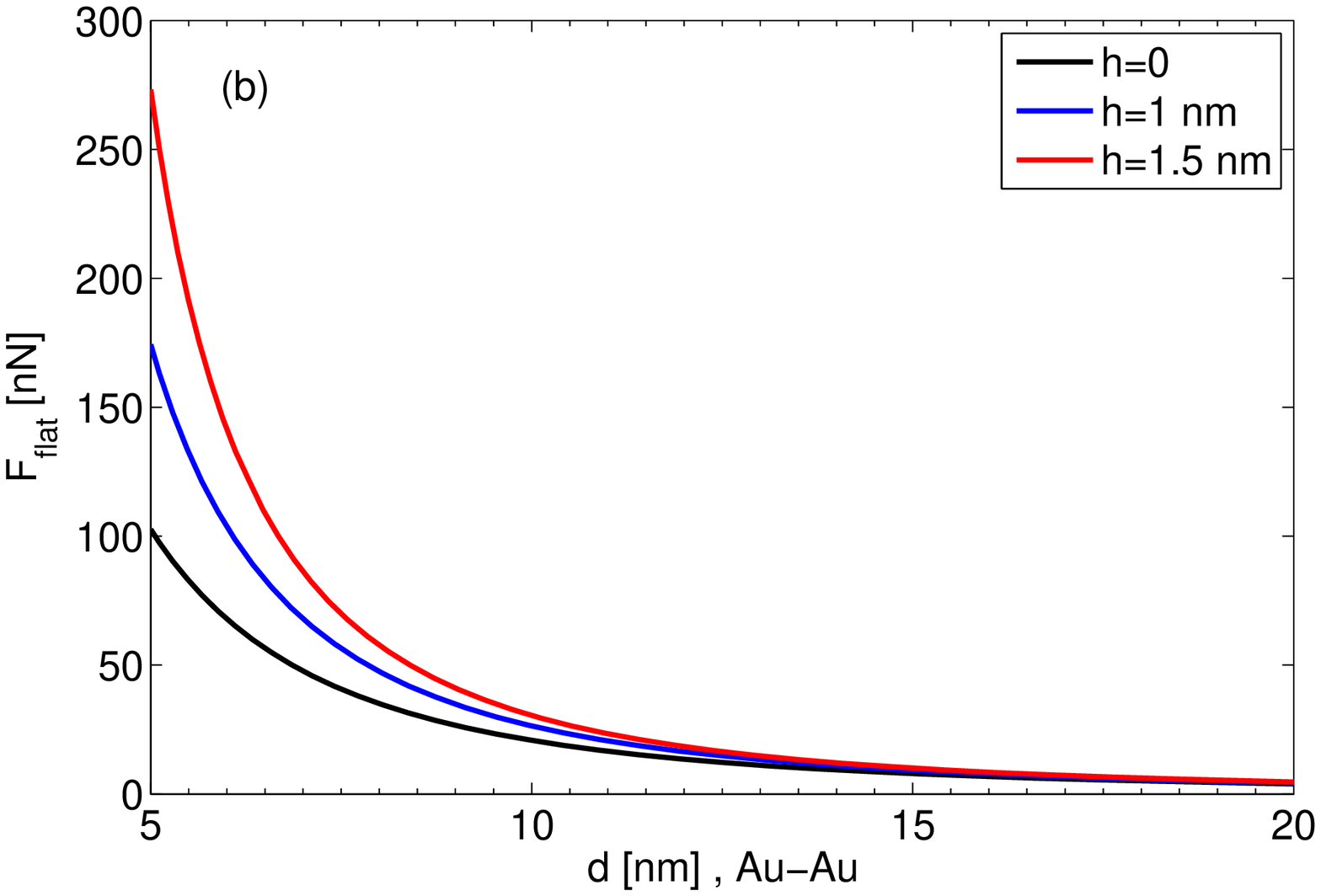}\newline
\caption{(a) Experimental data for the force vs distance (circles)
down to 13 nm separations, and the theoretical prediction without
water layer (red curve). Errors in the absolute separation are shown
for some points by the bars. The continuous black curve is the
prediction for continuous water layer of thickness $h=1.5$ nm. The
dashed black curve corresponds to the same water layer with 50\% of
voids ($f=0.5$). (b) Theoretical calculations using the Lifsthiz
theory for flat surfaces and continuous water layer films with
thicknesses $h=$0, 1, 1.5 nm  for small separations. } \label{fig12}
\end{center}
\end{figure}

Furthermore, formation of capillary bridges means that under ambient
conditions gold absorbs water, and as a result it is covered with an
ultra thin water layer. The experiments \cite{Zwo07,Zwo08c} suggest
that the thickness of this layer is in the nanometer range, 1-2 nm.
The natural questions one could ask is how thick the water layer is,
and what is the influence of this water layer on the dispersive
force \cite{Pal09}? At short separations, $d<20$ nm, these questions
become of crucial importance because they place doubts in our
understanding of the dispersive forces when experiments under
ambient conditions are compared with predictions of the Lifshitz
theory. Figure \ref{fig12} shows the Casimir force measured at short
distances together with theoretical calculations made for  gold with
or without a water layer on top. The errors are shown to arise
mainly from the experimental uncertainty in determining the
separation upon contact $d_0$ due to nanoscale surface roughness. We
can conclude that the experiment can exclude the water layer thicker
than 1.5 nm. Figure \ref{fig12}(b) shows that the effect of water
becomes very significant at separations below 10 nm, which were not
accessible in our measurements due to jump-into-contact. We
presented only the forces between flat surfaces because at these
small separations there is no a reliable way to estimate the
roughness correction.

\subsection{Distance upon contact}
\label{sec:3.3}

The absolute distance separating two bodies is a parameter of
principal importance for the determination of dispersive forces. It
becomes difficult to determine when the separation gap approaches
nanometer dimensions. This complication originates from the presence
of surface roughness, which manifests itself on the same scale. In
fact, when the bodies are brought into gentle contact they are still
separated by some distance $d_0$, which we call the distance upon
contact due to surface roughness. This distance has a special
significance for weak adhesion, which is mainly due to van der Waals
forces across an extensive noncontact area \cite{Del05}. It is
important for MEMS and NEMS as unremovable reason for stiction
\cite{Mab97}. In the modern precise measurements of the dispersive
forces $d_0$ is the main source of errors (see reviews
\cite{Lam05,Cap07}). This parameter is typically determined using
electrostatic calibration. The distance upon contact is usually
considerably larger than the rms roughness because it is defined by
the highest asperities. It is important to clear understand the
origin of $d_0$, its dependence on the lateral size $L$ of involved
surfaces, and possible uncertainties in its value \cite{Zwo09c}.
These are the questions we address in this subsection.

\subsubsection{Plate-plate contact}
\label{sec:3.3.1}

Two plates separated by the distance $d$ and having roughness
profiles $h_i(x,y)$ locally are separated by the distance $d(x,y)$
given by Eq. (\ref{loc_dist}) (see Fig. \ref{fig9}). Indeed, the
averaged local distance has to give $d$, $\left\langle
d(x,y)\right\rangle=d$. We can define the distance upon contact
$d_0$ as the largest distance $d=d_0$, for which $d(x,y)$ becomes
zero.

It is well known from contact mechanics \cite{Gre66} that the
contact of two elastic rough plates is equivalent to the contact of
a rough hard plate and an elastic flat plate with an effective
Young's modulus $E$ and a Poisson ratio $\nu$. Here we analyze the
contact in the limit of zero load when both bodies can be considered
as hard. This limit is realized when only weak adhesion is possible,
for which the dispersive forces are responsible. Strong adhesion due
to chemical bonding or due to capillary forces is not considered
here. This is not a principal restriction, but the case of strong
adhesion has to be analyzed separately.  Equation (\ref{loc_dist})
shows that the profile of the effective rough body is given by
\begin{equation}\label{comb}
    h(x,y)=h_1(x,y)+h_2(x,y).
\end{equation}
The latter means that $h(x,y)$ is given by the combined image of the
surfaces facing each other. If topography of the surfaces was
determined with AFM, we have to take the sum of these two images and
the combined image will have the size of the smallest image.

To determine $d_0$ we collected \cite{Zwo09c} high resolution
megascans (size $40\times 40\;\mu \textrm{m}^2$, lateral resolution
$4096\times 4096$ pixels) for gold films of different thicknesses
described before. The maximal area, which we have been able to scan
on the sphere, was $8\times 8\;\mu \textrm{m}^2$ ($2048\times 2048$
pixels). The images of 100 nm film, sphere, and 1600 nm film are
shown in Fig. \ref{fig1} (a), (b), and (c), respectively. Combining
two images and calculating from them the maximal peak height we can
find $d_0$ for a given size of the combined image. Of course, taking
the images of the same size every time we will get different value
of $d_0$ and averaging over a large number of images we will find
the averaged $d_0$ and possible rms deviations. This is quite
obvious. What is less obvious is that if we take images of different
size and will do the same procedure the result for $d_0$ will be
different.

\begin{figure}[ptb]
\begin{center}
\includegraphics[width=0.48\textwidth]{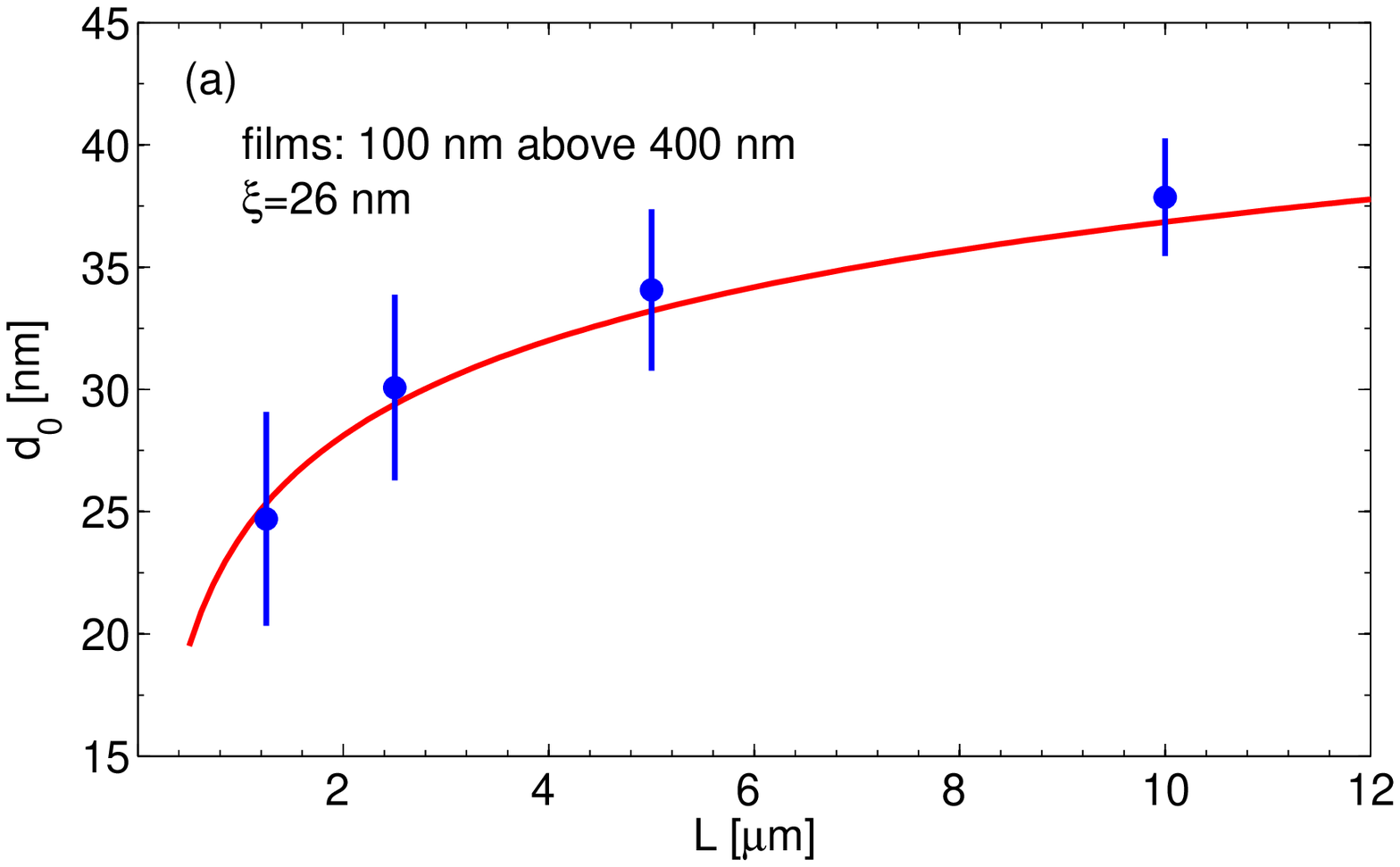}
\includegraphics[width=0.49\textwidth]{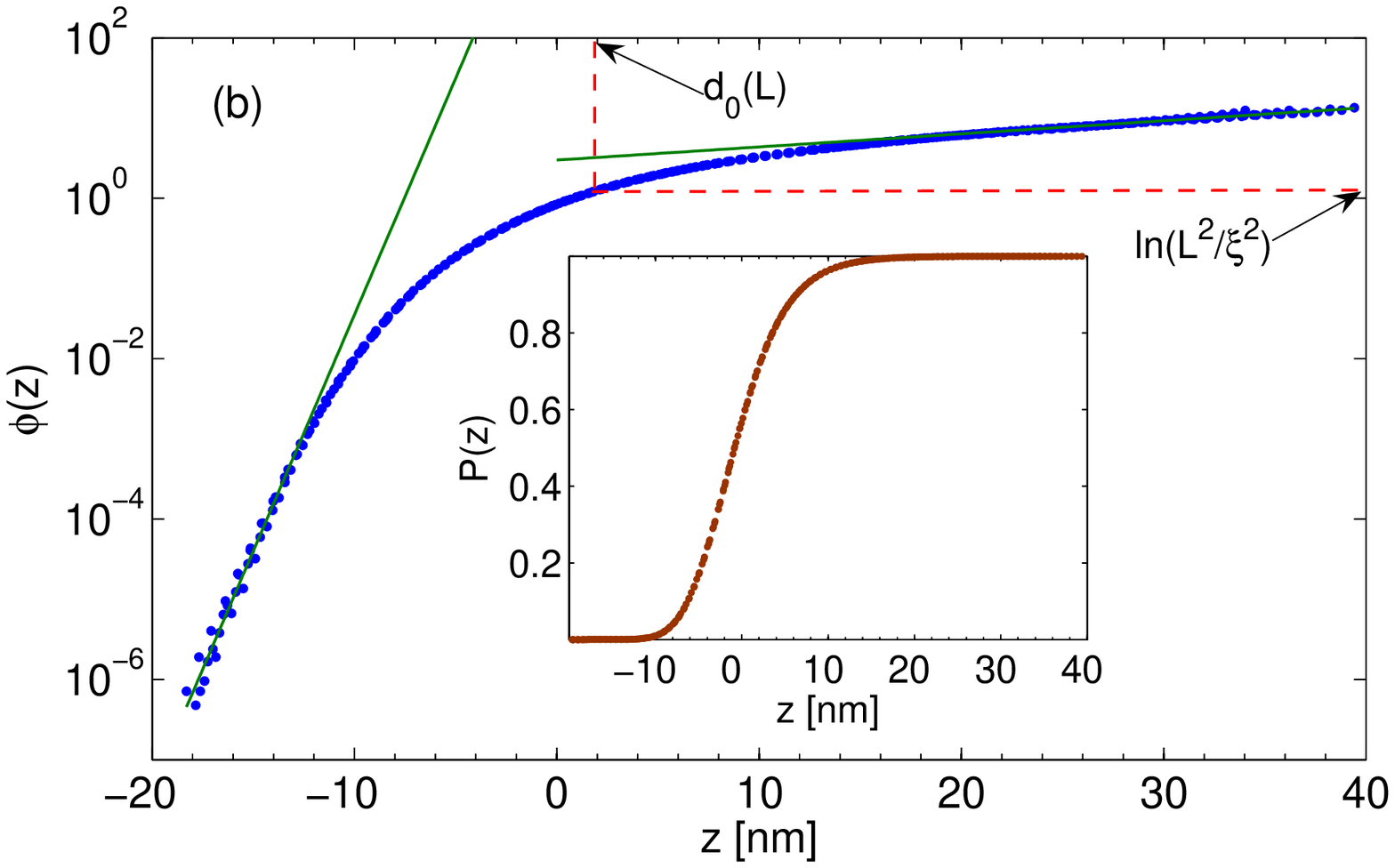}
\caption{(a) Distance upon contact as a function of the length
scale. Dots with the error bars are the values calculated from the
megascans. The solid curve is the theoretical expectation according
to Eq. (\ref{above_d0}). Note that $d_0$ is considerably larger than
$w$. (b) Statistics of the surface roughness. Four $10\times 10\;\mu
\textrm{m}^2$ images were used. The main graph shows the "phase" as
a function of $z$. The green lines show the best fits at large
positive and negative $z$. The dashed red lines demonstrate the
solution of Eq. (\ref{above_d0}). The inset shows the cumulative
distribution $P(z)$.} \label{fig13}
\end{center}
\end{figure}

Let $L_0$ be the size of the combined image. Then, in order to
obtain information on the scale $L=L_0/2^n$, we divide this image on
$2^n$ subimages. For each subimage we find the highest point of the
profile (local $d_0$), and average all these values. This procedure
gives us $d_0(L)$ and the corresponding statistical error. Megascans
are very convenient for this purpose otherwise one has to collect
many scans in different locations. For the 100 nm film above the 400
nm film the result of this procedure is shown in Fig.
\ref{fig13}(a). We took the maximum area to be $10\times 10\;\mu
\textrm{m}^2$. The figure clearly demonstrates the dependence of
$d_0$ on the scale $L$ although the errors appear to be significant.
The dependence of the rms roughness $w$ on the length scale $L$ is
absent in accordance with the expectations, while only the error
bars increase when $L$ is decreasing.

To understand the dependence $d_0(L)$ let us assume that the size
$L$ of the area of nominal contact is large in comparison with the
correlation length, $L\gg\xi$. It means that this area can be
divided into a large number $N^2=L^2/\xi^2$ of cells. The height of
each cell (asperity) can be considered as a random variable $h$. The
probability to find $h$ smaller than some value $z$ can be presented
in a general form
\begin{equation}\label{prob}
    P(z)=1-e^{-\phi(z)},
\end{equation}
where the "phase" $\phi(z)$ is a nonnegative and nondecreasing
function of $z$. Note that (\ref{prob}) is just a convenient way to
represent the data: instead of cumulative distributions $P(z)$ we
are using the phase $\phi(z)$, which is not so sharp function of
$z$.

For a given asperity the probability to find its height above $d_0$
is $1-P(d_0)$, then within the area of nominal contact one asperity
will be higher than $d_0$ if
\begin{equation}\label{above_d0}
    e^{-\phi(d_0)}\left({L^2}/{\xi^2}\right)=1\ \ \ \textrm{or}\ \ \
    \phi(d_0)=\ln\left({L^2}/{\xi^2}\right).
\end{equation}
This condition can be considered as an equation for the asperity
height because due to a sharp exponential behavior the height is
approximately equal to $d_0$. To solve (\ref{above_d0}) we have to
know the function $\phi(z)$, which can be found from the roughness
profile.

The cumulative distribution $P(z)$ can be extracted from combined
images by counting pixels with the height below $z$. Then the
"phase" is calculated as $\phi(z)=-\ln(1-P)$. The results are
presented in Fig. \ref{fig13}(b). The procedure of solving Eq.
(\ref{above_d0}) is shown schematically in Fig. \ref{fig13}(b) by
dashed red curves, and the solution itself is the red curve in Fig.
\ref{fig13}(a). It has to be mentioned that the normal distribution
fails to describe the data at large $z$. Other known distributions
cannot satisfactory describe the data for all $z$. Asymptotically at
large $|z|$ the data can be reasonably well fit with the generalized
extreme value Gumbel distributions (green lines in Fig.
\ref{fig13}(b)) \cite{Col01}:
\begin{equation}\label{Gumbel}
    \ln\phi(z)=\left\{\begin{array}{c}
                 -\alpha z,\ z\rightarrow -\infty \\
                 \beta z,\ z\rightarrow \infty
               \end{array}
               \right.
\end{equation}

The observed dependence $d_0(L)$ can be understood intuitively. The
probability to have one high asperity is exponentially small but the
number of asperities increases with the area of nominal contact.
Therefore, the larger the contact area, the higher probability to
find a high feature within this area. Scale dependence of $d_0$
shows that smaller areas getting into contact will be bound more
strongly than larger areas because upon the contact they will be
separated by smaller distances. This fact is important for weak
adhesion analysis.

\subsubsection{Sphere-plate contact}
\label{sec:3.3.2}

Most of the Casimir force experiments measure the force in the
sphere-plate configuration to avoid the problem with the plates
parallelism. Let us consider how the scale dependence of $d_0$
manifests itself in this case. Assuming that the sphere is large,
$R\gg d$, the local distance is
\begin{equation}\label{loc_sph_2}
    d(x,y)=d+\left(x^2+y^2\right)/2R-h(x,y),
\end{equation}
where $h(x,y)$ is the combined topography of the sphere and the
plate. As in the plate-plate case $d_0$ is the maximal $d$, for
which the local distance becomes zero. This definition gives
\begin{equation}\label{d0_sph}
    d_0=\max\limits_{x,y}\left[h(x,y)-\left(x^2+y^2\right)/2R\right].
\end{equation}
Now $d_0$ is a function of the sphere radius $R$, but, of course,
one can define the length scale $L_R$ corresponding to this radius
$R$.

As input data in Eq. (\ref{d0_sph}) we used the combined images of
the sphere and different plates. The origin ($x=0,\;y=0$) was chosen
randomly in different positions and then $d_0$ was calculated
according to (\ref{d0_sph}). We averaged $d_0$ found in 80 different
locations to get the values of $d_0^{im}$, which are collected in
Tab. \ref{tab3}. The same values can be determined theoretically
using $d_0(L)$ found between two plates (see Eq. (\ref{above_d0})
and Ref. \cite{Zwo09c}).

\begin{table}
\centering
\begin{tabular}{p{1.6cm}p{1.8cm}p{1.8cm}p{1.8cm}p{1.8cm}p{2cm}}
\hline\noalign{\smallskip}
 & $100\;nm$ & $200\;\textrm{nm}$ & $400\;\textrm{nm}$ & $800\;\textrm{nm}$ & $1600\;\textrm{nm}$ \\
\hline\hline $w$ & 3.8 & 4.2 & 6.0 & 7.5 & 10.1 \\
   $\xi$ & $26.1(3.8)$ & $28.8(3.7)$ & $34.4(4.7)$ & $30.6(2.4)$ & $42.0(5.5)$ \\
   $d_0^{im}$ & $12.8(2.2)$ & $15.9(2.7)$ & $24.5(4.8)$
   & $31.3(5.4)$ & $55.7(9.3)$ \\
   $d_0^{el}$ & $17.7(1.1)$ & $20.2(1.2)$ & $23.0(0.9)$
   & $34.5(1.7)$ & $50.8(1.3)$ \\
   \noalign{\smallskip}\hline\noalign{\smallskip}
\end{tabular}
\caption{The parameters characterizing the sphere-film systems (all
in nm) for $R=50\;\mu$m. The first three rows were determined from
combined images. The last row for $d_0^{el}$ gives the values of
$d_0$ determined electrostatically. The errors are indicared in
brackets. }\label{tab3}
\end{table}

The values of $d_0^{im}$ for rougher films are in agreement with
those found from the electrostatic calibration. However, for
smoother films (100 and 200 nm) $d_0^{im}$ and $d_0^{el}$ do not
agree with each other. This is most likely to be attributed to the
roughness on the sphere, which varies considerably locally. For
example, between those 80 $d_0^{im}$ found in different locations
5\% are in agreement with $d_0^{el}$ found from the electrostatic
calibration \cite{Zwo09c}. This is illustrated by the fact that when
the roughness of the plate dominates the discrepancy between
$d_0^{im}$ and $d_0^{el}$ disappears. Note that the standard
deviations for $d_0^{im}$ are larger than that for $d_0^{el}$. The
standard deviations in $d_0^{im}$ originate from place to place
variations of $d_0^{im}$. In the case of electrostatic determination
of $d_0$ statistical variation of $d_0$ from place to place is not
included in the errors of $d_0^{el}$. This explains why the errors
in $d_0^{el}$ are smaller.

\begin{figure}[ptb]
\begin{center}
\includegraphics[width=0.7\textwidth]{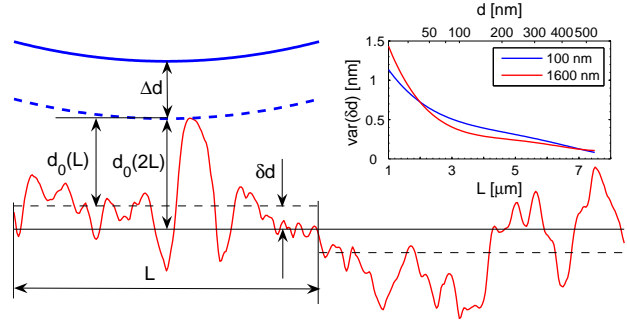}
\caption{Schematic explanation of additional uncertainty $\delta d$
in $d_0$ (see text). The sphere in two positions is shown by the
dashed (contact) and solid blue curves. The inset shows the variance
of $\delta d$ as a function of the scale $L$ or separation $d$. }
\label{fig14}
\end{center}
\end{figure}

Consider the experimental situation when the dispersive force is
measured in the sphere-plate configuration. The system under
consideration is equivalent to a smooth sphere above a combined
rough profile $h(x,y)$. The position of the average plane depends on
the area of averaging $L^2$ especially for small scales $L$. The
profile shown in Fig. \ref{fig14} demonstrates different mean values
in the left and right segments shown by the dashed black lines. Both
of these values deviate from the middle line for the scale $2L$
(solid black line). The true average plane is defined for
$L\rightarrow\infty$.

Position of the average plane define the absolute separation of the
bodies. It has to be stressed that the electrostatic and Casimir
interactions "see" different areas on the plate. This is due to
different dependence on $d$ and quite often the electrostatic
calibration is performed at larger separations than measurement of
the Casimir force. The size $L$ of the interaction area is
determined by the relation $L^2=\alpha\pi Rd$, where $\alpha=2$ for
the electrostatic and $\alpha=2/3$ for pure Casimir force.
Therefore, these two interactions can "see" different positions of
the average planes. It introduces an additional uncertainty $\delta
d$ in the absolute separation \cite{Zwo09c}. For a fixed $L$ this
uncertainty is a random variable distributed roughly normally around
$\delta d=0$. However, it has to be stressed that $\delta d$
manifests itself not as a statistical error but rather as a kind of
a systematic error. This is because at a given lateral position of
the sphere this uncertainty takes a fixed value. The variance of
$\delta d$ is defined by the roughness statistics. It was calculated
from the images and shown as inset in Fig. \ref{fig14}. One has to
remember that with a probability of 30\% the value of $\delta d$ can
be larger than that shown in Fig. \ref{fig14}.

\section{Conclusions}
\label{sec:4}

In this chapter we considered the Casimir force between realistic
materials containing defects, which influence the optical properties
of interacting materials, and having surface roughness, which
contributes to the force.

It was demonstrated that the gold films prepared in different
conditions have dielectric functions, which differ from sample to
sample, and this difference cannot be ignored in the calculation of
the Casimir force aimed at precision better than 10\%. The main
conclusion is that for metals one has to measure the dielectric
function of used materials in a wide range of frequencies, where far
and mid IR are especially important. Precise knowledge of the
dielectric functions is also important for low dielectric materials.
In this case significant sensitivity of the force to the dielectric
functions is realized nearby the attractive-to-repulsive transition
in solid-liquid-solid systems.

The roughness correction to the Casimir force can be reliably
calculated if rms roughness $w$ is small in comparison with the
separation, $w\ll d$, when one can apply the perturbation theory. In
the experiments at short separations this condition can be violated.
The current situation with the theory is that there is no direct
method to calculate the force between rough bodies when $d\sim  w$
except using rather complicated numerical calculations.

We gave also a detailed analysis of the distance upon contact,
$d_0$, which is an important parameter in Casimir physics. Analysis
of AFM scans demonstrated that $d_0$ is always a few times larger
than the rms roughness. Moreover, $d_0$ is a function of the size
$L$ of the nominal area of contact. This dependence is important for
weak adhesion, which is due to van der Waals forces across an
extensive noncontact area. Uncertainty in $d_0$ is the main source
of errors in the Casimir force measurements. We demonstrated here
that there is an additional indefiniteness in $d_0$, which cannot be
excluded by the electrostatic calibration. It becomes very important
for small areas of interaction. Also, this indefiniteness has to be
taken into account if one compares two independent experiments.

\begin{acknowledgement}
The research was carried out under Project No. MC3.05242 in the
framework of the Strategic Research Programme of the Materials
Innovation Institute M2i (the former Netherlands Institute for
Metals Research NIMR) The authors benefited from exchange of ideas
by the ESF Research Network CASIMIR.
\end{acknowledgement}

\end{document}